\pdfoutput=1
\documentclass[a4paper,UKenglish]{lipics-v2016}
 
\usepackage{amssymb}
\usepackage[ruled, vlined,linesnumbered]{algorithm2e}
\usepackage{url}
\usepackage{amsthm}
\usepackage{microtype}
\usepackage{stmaryrd}  
\usepackage{mathtools}  
\usepackage{tikz}  
\usetikzlibrary{matrix,arrows}

\newcommand{\sem}[1]{\llbracket{#1}\rrbracket}
\newcommand{\Const}{\mbox{\sf Const}}
\newcommand{\Null}{\mbox{\sf Null}}
\newtheorem{proposition}[theorem]{Proposition}
\title{Temporal Data Exchange}

\author[1]{Ladan Golshanara}
\author[1]{Jan Chomicki}
\affil[1]{State University of New York at Buffalo, NY, USA\\
  \texttt{ladangol@buffalo.edu, chomicki@buffalo.edu }}


\begin{document}

\maketitle

\begin{abstract}
   Data exchange is the problem of transforming data
that is structured under a source schema into data structured under
another schema, called the target schema,
so that 
both the source and target data satisfy the relationship between the schemas.
Many applications 
such as planning, scheduling, medical 
and fraud detection systems, require data exchange
 in the context of temporal data.
Even though the formal framework of data exchange for
relational database systems is well-established, it does not immediately
carry over to the settings of temporal data,
which necessitates reasoning over unbounded periods of time.

In this work, we study data exchange for temporal data. We first motivate
the need for two views of temporal data: the concrete view, which 
depicts how temporal data is compactly represented and on which the implementations are based, and the abstract view, which
defines the semantics of temporal data as a sequence of snapshots. 
We first extend the chase procedure for the abstract view to have a conceptual basis for the data exchange for temporal databases. Considering non-temporal source-to-target tuple generating dependencies and equality generating dependencies, the chase algorithm can be applied on each snapshot independently. Then we define a chase procedure (called {\em c-chase}) on concrete instances and show the result of c-chase on a concrete instance is semantically aligned with the result of chase on the corresponding abstract instance. 
In order to interpret intervals as constants while checking if a dependency or a query is satisfied by a concrete database, we will {\em normalize} the instance with respect to the dependency or the query. To obtain the semantic alignment, the nulls (which are introduced by data exchange and model incompleteness) in the concrete view are annotated with temporal information. 
Furthermore, we show that the result of the concrete chase provides a foundation for query answering. We define na\"ive evaluation on the result of the c-chase and show it produces certain answers.
\end{abstract}

\section{Introduction}
\label{intro}
{\em Temporal data} refers to historical data or data that is 
dated.  Temporal data is
needed by many organizations and individuals to support audit trails. With temporal data one can represent when a fact is true and for how long \cite{Dong15}.  The temporality of
facts is also critical in diverse domains, from medical diagnosis to
assessing the changing business conditions of companies to taxi and bicycle rides \cite{SNG12}. To support temporal database applications suitable database features  were
recently added to the SQL:2011 standard~\cite{Kulkarni12}, and adopted by major database management systems such as DB2,
Oracle, and Teradata. The need for data integration and data exchange in the context of temporal data is discussed in \cite{RothT13, AlexeRT14, DongKT16}.

{\em Temporal databases} provide a uniform and systematic way of dealing with historical data \cite{Chomicki05}.  Prior work on temporal databases~\cite{Koubarakis94b, Toman96, Chomicki05}
has provided two views of temporal data: the {\em abstract temporal view} (or {\em abstract view} in short) and the {\em concrete temporal view} (or {\em concrete view} in short). Abstract view provides representation-independent meaning of a temporal database while concrete view provides a finite representation of temporal data. Conceptually, we can associate to each time point $\ell$ the state $db_{\ell}$ of a database at the time point $\ell$. Thus, a temporal database in the abstract view is a sequence of states ({\em snapshots}). The domain of time points is a totally ordered set which is isomorphic to non-negative integers $\mathbb{N}_0$. For example, consider a database schema $E(name, company)$ and the fact that Ada worked in IBM between 2010 and 2013. In the abstract view the snapshots of $E$ associated with the time points 2010 to 2013 contain a fact $E(Ada, IBM)$.
Due to repetitive data in consecutive snapshots, storing information in the abstract view is not practical and is meant only to provide the semantics for the concrete view. In the concrete view, temporal data is summarized in a single database instance in which data is time-stamped with a {\em time interval}\footnote{We assume time intervals have the format $[s,e)$, where $s,e \in \mathbb{N}_0$ and $e$ can be $\infty$.} that indicates when the fact is true. The concrete view is an extension of the relational model where each relation in a database is augmented with a temporal attribute which takes time intervals as values. 
 For example, in the concrete view the information above about $Ada$ is usually represented as $E(Ada, IBM, [2010, 2014))$ where
[2010, 2014) denotes the years 
2010, 2011, 2012, and 2013. The fact that 
Ada has worked in Intel since then can be represented
as E(Ada, Intel, [2014, $\infty$)). An infinite time interval, such as [2014, $\infty$), is a useful abstraction when the endpoint is not provided.

 Data exchange~\cite{FKMP05} refers to the problem of translating data that conforms to
one schema (called the {\em source schema $R_S$}) into data that conforms to 
another schema (called the {\em target schema $R_T$}), given a specification
of the relationship between the two schemas. This relationship is 
specified by means of a {\em schema mapping} consisting of a set of {\em source-to-target tuple generating
dependencies (s-t tgds)} and a set of {\em tuple generating dependencies (tgds)} and {\em equality generating dependencies (egds)} on the target schema.
Given a schema mapping and a source instance $I$, the goal of data exchange is to materialize
a target instance $J$ that satisfies the specification (i.e. $(I, J)$ satisfies s-t tgds and $J$ satisfies tgds and egds). Such an instance $J$ is called a {\em solution} for $I$ w.r.t. the given schema mapping.
For a given source instance, there may be no solution since
there may not exist a target instance that satisfies the specification.
On the other hand, there may be many 
solutions. 
It was shown in~\cite{FKMP05} that among all solutions for a given
source instance, the {\em universal solutions} are the preferred solutions
because they are the most general. In~\cite{FKMP05}, the {\em chase procedure} is used to find a universal solution for a given source instance with respect to a schema mapping. Universal solutions can be used to determine the {\em certain answers} to unions of conjunctive queries posed over a target schema. Certain answers to a query $q$ are the tuples that are in the answer of $q$ in any solution for a source instance w.r.t. a schema mapping.

In this paper, we study the challenges that arise when we consider temporal data in the framework of data exchange. We examine the most basic case where the s-t tgds and egds are non-temporal, that is they refer to single snapshots. In such a case, each snapshot in an abstract instance is treated independently from the past or future snapshots. We do not consider tgds to avoid dealing with non-termination issues \cite{GrecoST15, GrahneO18} of the chase procedure which are orthogonal to temporal database issues. After defining the semantics of data exchange on the abstract instances, we define data exchange on concrete instances. First a universally quantified variable $t$ is added to each s-t tgd and egd to match the schema of concrete databases, however these dependencies are still implicitly non-temporal. In order to interpret time intervals as constants while checking if a conjunctive formula is satisfied by a concrete instance, we will {\em normalize} the concrete instance such that the time intervals behave as constants w.r.t. a set of conjunctive formulas. 

We discuss how to handle unknown values in concrete target instances. A concrete fact spans multiple consecutive snapshots in the abstract view. Since the chase in the abstract view produces fresh labeled nulls in each snapshot independently from the past or future snapshots, the chase for the concrete view needs to produce an unknown value that corresponds to distinct labeled nulls in consecutive snapshots in the abstract view.
Thus, we introduce {\em interval-annotated nulls} to represent unknown values generated as a result of data exchange. These nulls are annotated with the time interval of the concrete facts they occur in.
 For example, $N^{[s,e)}$ is an interval-annotated null in a concrete fact with the time interval $[s,e)$. An interval-annotated null is a representation of a sequence of labeled nulls. Consider a concrete fact $Emp(Ada, IBM, N^{[8, \infty)}, [8,\infty))$. The interval-annotated null $N^{[8,\infty)}$ represents the sequence of labeled nulls $\langle N_8, N_9 , \ldots \rangle$. In the abstract view, the snapshot $db_8$ contains the fact $Emp(Ada,\ IBM,\ N_{8})$, the snapshot $db_9$ contains the fact $\allowbreak Emp(Ada,\ IBM,\ N_{9})$ and so on. 

{\bf Contributions~}
Our main contribution of this paper is the formalization and study of 
the framework of data exchange on temporal data. We extend the data exchange on the abstract view which provides the semantics of temporal data exchange. We show the result of a successful chase on the abstract view is a universal solution. Then we propose a concrete chase for the concrete view.
We show the correctness of the concrete chase by showing that its result has correct semantics. 
This result is important because it enables one to implement data exchange on concrete temporal data  
with semantics corresponding to 
the abstract view. We introduce interval-annotated nulls to represent the unknown values generated as a result of data exchange on the concrete view.
Finally, we define na\"ive evaluation on the concrete instances and show that the result of concrete chase can be used to find certain answers in the corresponding abstract view.

The rest of this paper is organized as follows. Section \ref{sec:background} recalls some background on temporal database and data exchange and introduces some of the notations we use in this text. Section \ref{sec:ADX} discusses the abstract data exchange. Section \ref{sec:CDX} introduces concrete chase. In this section we show the result of a concrete chase has correct semantics. Section \ref{sec:QA} studies query answering in the context of temporal data exchange. Section \ref{sec:relatedwork} discusses related work and Section \ref{sec:conclusion} concludes the paper.

\section{Background}\label{sec:background}
 We assume a fixed database schema $\mathbf{R}$. An abstract temporal database instance ({\em abstract instance} for short) $I_a$ is an infinite sequence of snapshots $\langle db_0, db_1, db_2, \ldots \rangle$. Each $db_{\ell}$ ($\ell \in \mathbb{N}_0$) is a relational database instance over $\mathbf{R}$. Each snapshot in an abstract source instance consists of facts: expressions of the form $R(a_1, \ldots, a_n)$ where $R$ is an n-ary relation name $R(A_1, \ldots, A_n)$ in the schema, $A_1 \ldots A_n$ are (data) attributes and $a_1, ..., a_n$ are constants. Sometimes we denote a vector of constants by $\mathbf{a}$. In the abstract target instance the relations might contain labeled nulls as well. We denote by $\mathbf{N}$ a vector of labeled nulls, and by $\Null(db)$ the labeled nulls that occur in the relational database instance $db$.

We assume abstract instances satisfy the {\em finite change condition} which indicates that there exists $m \in \mathbb{N}_0$ such that $db_m = db_{m+1} = \ldots$ \cite{ChomickiW16}. 

The finite change condition intuitively means from some time point on the state of the temporal database remains unchanged. The practical consequence is that an abstract instance can be represented by a finite concrete instance by time stamping the fact with a time interval $[s,e)$ or $[s, \infty)$ where $s,e \in \mathbb{N}_0$.

If $\mathbf{R}$ is a database schema, we denote by $\mathbf{R}^+$ the corresponding concrete database schema such that for each n-ary relation $R(A_1,\ldots, A_n)$ in $\mathbf{R}$ there is a $(n+1)$-ary concrete relation, denoted by $R^+(A_1,\ldots, A_n, T)$ in $\mathbf{R}^+$ where $T$ is the temporal attribute and $A_1, \ldots A_n$ are data attributes.
The domain of the temporal attribute consists of time intervals of the form $[s,e)$ or $[s,\infty)$, where $s, e \in \mathbb{N}_0$. 

We use the symbols $I_c, J_c, I'_c, J'_c$ (resp. $I_a, J_a, I'_a, J'_a$) to refer to concrete instances (resp. abstract instances).
If a concrete (resp. abstract) database instance does not contain unknown information (nulls) we call it a  {\em complete} concrete (resp. abstract) instance. If $I_c$ is a complete concrete instance, then we denote by $\sem{I_c}$ the abstract database instance that $I_c$ represents \cite{ChomickiW16}, that is:

$$\sem{I_c} = \langle db_0, db_1, \ldots \rangle$$
such that for all $\ell \in \mathbb{N}_0$, 
$$db_{\ell} = \{\ R(\mathbf{a})\  |\  \exists s. \exists e.\  \ R^+(\mathbf{a}, [s,e)) \in I_c \text { and  } s \leq \ell < e\}$$

A concrete instance is {\em coalesced} if the facts with identical data attribute values have disjoint (i.e. non- overlapping) or non-adjacent time intervals \cite{ Bohlen96, Chomicki05}. Two intervals $[s,e)$, $[s',e')$ are adjacent if $s' = e$ or $s=e'$. Any abstract database can be represented by a unique coalesced concrete database. We assume the concrete instances are coalesced in this paper. 

As in the standard data exchange paper \cite{FKMP05} we assume abstract source instances contain only constants (and in case of concrete instances, constants and time intervals). Thus, the abstract and concrete source instances are complete.

A non-temporal s-t tgd is of the form $$\sigma_{st}: \forall \mathbf{x} \  \phi(\mathbf{x}) \rightarrow \exists \mathbf{y} \psi(\mathbf{x}, \mathbf{y}) $$ 
and an egd is of the form
$$\sigma_{eg}: \forall \mathbf{x} \ \phi(\mathbf{x}) \rightarrow x_1 = x_2$$
where $\mathbf{x}$ and $\mathbf{y}$ are vectors of variables and $x_1$ and $x_2$ are variables in $\mathbf{x}$. In the rest of the paper we will usually drop the universally quantified variables.

The s-t tgds and the egds on concrete schemas are augmented with a universally quantified variable $t$ in each atom in the left-hand-side (lhs for short) and right-hand-side (rhs for short) of the dependency:
$$\sigma^+_{st}: \forall \mathbf{x},t \ \phi(\mathbf{x},t) \rightarrow \exists \mathbf{y} \ \psi(\mathbf{x}, \mathbf{y}, t)$$
$$\sigma^+_{eg}: \forall \mathbf{x},t \  \phi(\mathbf{x}, t) \rightarrow x_1 = x_2$$

The domain ({\em sort}) of variable $t$ is time intervals.
A {\em data exchange setting} is a quadruple $\mathcal{M}=(R_S, R_T, \Sigma_{st}, \Sigma_{eg})$ where $R_S$ and $R_T$ are the source and target schemas, respectively; $\Sigma_{st}$ is a set of s-t tgds and $\Sigma_{eg}$ is a set of egds. The source and the target schemas are disjoint. The corresponding setting for concrete databases is $\mathcal{M}^+=(R^+_S, R^+_T, \Sigma^+_{st}, \Sigma^+_{eg})$.

In~\cite{FKMP05}, the {\em chase procedure} is used to find a universal solution for a data exchange setting $\mathcal{M}$.
A solution is universal if it has {\em homomorphisms} to every other solution. A {\em homomorphism} $h$ from a relational instance $J_1$ to another instance $J_2$, denoted by $h: J_1 \rightarrow J_2$, is a function from the constants and labeled nulls in $J_1$ to constants and labeled nulls in $J_2$ such that:
\begin{itemize}
    \item $h(a) = a$, where $a$ is a constant in $J_1$
    \item $h(N_0) = v$, where $N_0$ is a labeled null in $J_1$ and $v$ is either a constant or a labeled null.
    \item for every $R(v_1, ..., v_n) \in J_1$, $R(h(v_1), ..., h(v_n))$ is in $J_2$.
\end{itemize}
A homomorphism $h$ is also used for a mapping from a dependency (such as an s-t tgd or an egd) to an instance $I$ such that for every atom $R(\mathbf{x})$ in the dependency $R(h(\mathbf{x}))$ is a fact in $I$. 

 The standard chase modifies an instance by a sequence of {\em chase steps} until all dependencies are satisfied. A chase step is {\em fired} by a homomorphism and a dependency. If the dependency is a tgd, a chase step  generates new facts in the target instance. Also, fresh labeled nulls are generated at each tgd chase step for each existentially quantified variable. If the dependency is an egd, then the chase step might be a successful or not. If the chase step is successful, then some labeled nulls in facts are replaced by other labeled nulls or constants. If one constant is equated to another constant, the chase step fails. For formal definition of the chase procedure refer to \cite{FKMP05}. 
 
 \section{Abstract data exchange}\label{sec:ADX}
In this section we extend the standard chase procedure to abstract instances. The s-t tgds and egds we consider are the ones introduced in \cite{FKMP05} which are over relational databases. 
Consider a data exchange setting $\mathcal{M} = (R_S, R_T, \Sigma_{st}, \Sigma_{eg})$. Since the s-t tgds and egds are non-temporal, in order to apply the chase procedure on an abstract source instance $I_a$ w.r.t. $\mathcal{M}$, we apply the chase procedure to each snapshot independently, that is 
$$chase(I_a, \mathcal{M}) = \langle chase(db_0, \mathcal{M}), chase(db_1, \mathcal{M}), \ldots \rangle$$
The fresh labeled nulls that are produced in a snapshot are distinct from the labeled nulls produced in the other snapshots. Otherwise, it means that an unknown value appears in different snapshots which is not intended by non-temporal s-t tgds and egds. 
 
If the result of at least one of the chase procedures on a snapshot is a failure, then the result of $chase(I_a, \mathcal{M})$ is a failure.

\begin{example}\label{ex:runex}
Consider a source schema with two relations $E(name, company)$ and $S(name, salary)$. Some snapshots of the abstract view of the temporal database are shown in Figure \ref{fig: snapshotsrc}. 

\begin{figure}
    \centering
     \begin{tabular}{|c|c|}  
  \multicolumn{1}{l}{$I_a$} & \multicolumn{1}{r}{} \\ \hline
  $2012$ & \{E(Ada, IBM)\} \\ \hline
  $2013$ & \{E(Ada, IBM), S(Ada, 18k), E(Bob, IBM)\}\\ \hline
  $2014$ & \{E(Ada, Google), S(Ada, 18k), E(Bob, IBM)\}\\ \hline
  $2015$ & \{E(Ada, Google), S(Ada, 18k), E(Bob, IBM), S(Bob, 13k)\}\\ \hline
  $\ldots$ & $\ldots$ \\ \hline
  $2018$ & \{E(Ada, Google), S(Ada, 18k), S(Bob, 13k)\} \\ \hline
  $\ldots$ & $\ldots$ \\ \hline
  \end{tabular}
  \caption{\label{fig: snapshotsrc}Some snapshots in the abstract view of a temporal instance.}
\end{figure}
We have the following non-temporal s-t tgds:
$$\forall n,c \  E(n,c) \rightarrow \exists s Emp(n,c,s)$$
$$\forall  n,c,s \ E(n,c) \wedge S(n,s) \rightarrow Emp(n,c,s)$$
and the following egd:
$$\forall n, c, s, s' \ Emp(n,c,s) \wedge Emp(n,c,s') \rightarrow s= s'$$
\end{example}

A target abstract instance $J_a$ is a {\em solution} for a source instance $I_a$ w.r.t. a data exchange setting $\mathcal{M}$ if each snapshot $db_{\ell}$ ($\ell \in \mathbb{N}_0$) in $(I_a, J_a)$ is a solution, that is $db_{\ell} \models 
(\Sigma_{st} \cup \Sigma_{eg})$, where $\models$ shows the dependencies in $\Sigma_{st} \cup \Sigma_{eg}$ are satisfied by $db_{\ell}$.

Consider two abstract instances $I_a = \langle db_0, db_1, \ldots \rangle$ and $I'_a = \langle db'_0, db'_1, \ldots \rangle$. There exists a homomorphism $h$ from $I_a$ to $I'_a$ (i.e. $h: I_a \mapsto I'_a$) if:
 \begin{enumerate}
     \item There is a homomorphism $h_{\ell}: db_{\ell} \mapsto db'_{\ell}$, $\ell \in \mathbb{N}_0$
     \item
     \begin{equation*}
         \begin{split}
             & \forall i, j \in \mathbb{N}_0, \ i\neq j \text{ such that } \\
             &  h_i: db_i \mapsto db'_i \text{ and } h_j: db_j \mapsto db'_j , \\
            & \forall \ell \in \mathbb{N}_0.\  \forall N \in \Null(db_{\ell}),   \  \  h_i(N) = h_j(N) 
         \end{split}
     \end{equation*}
  
 \end{enumerate}
Example \ref{ex:unionhomo} shows why the second condition above is necessary.
 \begin{example}\label{ex:unionhomo}
Consider the target schema $Emp(name, company, salary)$. Two instances of the target schema are shown in Figure \ref{fig:unionhomo}.
\begin{figure}
    \centering
     \begin{tabular}{|c|c|}  
  \multicolumn{1}{l}{$J_1$} & \multicolumn{1}{r}{} \\ \hline
  $db_0$ & Emp(Ada, IBM, $N$) \\ \hline
  $db_1$ & Emp(Ada, IBM, $N$)\\ \hline
  \end{tabular}
  \quad
  \quad
  \begin{tabular}{|c|c|}  
  \multicolumn{1}{l}{$J_2$} & \multicolumn{1}{r}{} \\ \hline
  $db'_0$ &Emp(Ada, IBM, $M_1$) \\ \hline
  $db'_1$ & Emp(Ada, IBM, $M_2$)\\ \hline
  \end{tabular} 
    \caption{Two abstract instances with nulls}
    \label{fig:unionhomo}
\end{figure}
In the instance $J_1$ the nulls in two consecutive snapshots are the same, representing one unknown value. 
Though from each snapshot in $J_1$ there is a homomorphism to the corresponding snapshot in $J_2$, that is $h_1: db_0 \mapsto db'_0$ and $h_2: db_1 \mapsto db'_1$, they do not agree on mapping $N$, that is $h_1(N) \neq h_2(N)$. 
\end{example}

 In the Example \ref{ex:unionhomo}, there is a homomorphism from the instance $J_2$ to $J_1$, but there is no homomorphism from $J_1$ to $J_2$.
 
\begin{definition}\label{def:unisol}{\em Universal solution:} 
 A target instance $J_a= \langle db_0, db_1, \ldots \rangle$ is a {\em universal solution} for $I_a$ w.r.t. a data exchange setting $\mathcal{M}$ if $J_a$ is a solution and for an arbitrary solution $J'_a=\langle db'_0, db'_1, \ldots \rangle$ for $I_a$ w.r.t. $\mathcal{M}$, there exists a homomorphism $h: J_a \mapsto J'_a$.
\end{definition}

\begin{proposition}\label{prop:uniabstract}
Let $\mathcal{M}= (R_S, R_T, \Sigma_{st}, \Sigma_{eg})$ be a data exchange setting. Let $I_a$ be an abstract source instance. 
\begin{enumerate}
    \item The result of a successful $chase(I_a, \mathcal{M})$ is a universal solution.
    \item If the result of $chase(I_a, \mathcal{M})$ is failure then there is no solution.
\end{enumerate} 

\end{proposition} 
\begin{proof}
\noindent\textbf{{\em Part 1:}}
Let $J_a = \langle db_0, db_1, \ldots \rangle$ be the target instance obtained by chase. 
Let $J'_a = \langle db'_0, db'_1, \ldots \rangle$ be any solution for $I_a$ with respect to $\mathcal{M}$. Based on Theorem 3.3 in \cite{FKMP05}, the result of a successful chase on each snapshot is a universal solution, meaning that there is a homomorphism $h_{\ell}$ from each snapshot $db_{\ell}$ in $J_a$ to the corresponding snapshot $db'_{\ell}$ in $J'_a$, $\ell \in \mathbb{N}_0$. Each of the homomorphisms defined from a snapshot in $J_a$ to the corresponding snapshot in $J'_a$ is identity on constants. Now we need to show that these homomorphisms meet the second condition in the Definition \ref{def:unisol}. The labeled nulls that are produced by the chase procedure in each snapshot in $J_a$ are different from the labeled nulls in other snapshots (by definition), that is $~\forall i \in \mathbb{N}_0. ~\forall j \in \mathbb{N}_0. ~ \  (Null(db_{i}) \cap Null(db_{j})) = \emptyset$. Therefore, the homomorphisms $h_0, h_1, \ldots $ can be extended in the following way:

\[h'_{\ell}(N) = 
\left\{
	\begin{array}{ll}
	h_0(N)  \mbox{ if } N \in \Null(db_0) \\ 
	h_1(N)  \mbox{ if } N \in \Null(db_1) \\ 
	\ldots \\
	h_{\ell}(N)  \mbox{ if } N \in \Null(db_{\ell}) \\
	\ldots \\
	\end{array}
\right.\]
Hence, $J_a$ is a universal solution.

\noindent\textbf{{\em Part 2:}} Let $I_a = \langle db''_0, db''_1, \ldots \rangle$. If the result of $chase(I_a, \mathcal{M})$ is a failure, then it means for some $\ell \in \mathbb{N}_0$ the result of $chase(db''_\ell, \mathcal{M})$ is a failure, where $db''_{\ell}$ is a snapshot in $I_a$. Based on the Theorem 3.3 in paper \cite{FKMP05}, there is no solution for the snapshot $db''_{\ell}$. Therefore there is no target instance $J_a$ such that $(I_a, J_a) \models (\Sigma_{st} \cup \Sigma_{eg})$ (because  $\Sigma_{eg}$ is not satisfied in the $\ell^{th}$ snapshot of $(I_a, J_a)$).   
\end{proof}
\begin{example}\label{ex:s-teg}
The result of applying chase on each snapshot of $I_a$ from Figure \ref{fig: snapshotsrc} is shown in Figure \ref{fig: snapshottar}.
\end{example}
\begin{figure}
    \centering
     \begin{tabular}{|c|c|}  
  \multicolumn{1}{l}{$J_a$} & \multicolumn{1}{r}{} \\ \hline
  $2012$ & \{Emp(Ada, IBM, $N$)\} \\ \hline
  $2013$ & \{Emp(Ada, IBM, 18k), Emp(Bob, IBM, $N'$)\}\\ \hline
  $2014$ & \{Emp(Ada, Google, 18k), Emp(Bob, IBM, $M$)\}\\ \hline
  $2015$ & \{Emp(Ada, Google, 18k), Emp(Bob, IBM, 13k)\}\\ \hline
  $\ldots$ & $\ldots$ \\ \hline
  $2018$ & \{Emp(Ada, Google, 18k)\} \\ \hline
  $\ldots$ & $\ldots$ \\ \hline
  \end{tabular}
  \caption{\label{fig: snapshottar} Some snapshots of the abstract view of the result of the chase procedure w.r.t. to the data exchange setting discussed in the Example \ref{ex:runex}}
\end{figure}
\section{Concrete data exchange}\label{sec:CDX}
In this section, we define a chase algorithm called {\em c-chase} for a data exchange setting $\mathcal{M}^+ = (R^+_S, R^+_T, \Sigma^+_{st}, \Sigma^+_{eg})$ and a concrete source instance. Note that although each dependency in $\Sigma^+_{st}$ is augmented with universally quantified variable $t$, these s-t tgds and egds are implicitly non-temporal because they lack the expressive power to express the temporal phenomena such as an event happened {\em before} another event.
\begin{example}\label{ex:conrunex}
The concrete view of the temporal database shown in Figure \ref{fig: snapshotsrc} is shown in Figure \ref{fig:coninstance}.
\begin{figure}
    \centering

  \begin{tabular}{|c|c|c|}  
  \multicolumn{1}{l}{$E^+$} & \multicolumn{1}{r}{} & \multicolumn{1}{r}{}\\ \hline
  Name & Company & Time \\ \hline
   Ada & IBM &  [2012, 2014) \\ \hline
   Ada & Google & [2014, $\infty$) \\ \hline
   Bob & IBM & [2013, 2018) \\ \hline
  \end{tabular}
  \quad 
  \quad
  \begin{tabular}{|c|c|c|}
    \multicolumn{1}{l}{$S^+$} & \multicolumn{1}{r}{} & \multicolumn{1}{r}{}\\ \hline
       Name&Salary& Time  \\ \hline
       Ada&18k&[2013, $\infty$) \\ \hline
       Bob & 13k & [2015, $\infty$) \\ \hline
  \end{tabular}
    \caption{A concrete source instance $I_c$}
    \label{fig:coninstance}
\end{figure}
The s-t tgds and the egd are as follows:
$$\sigma_1^+: \forall n,c,t \  E^+(n,c,t) \rightarrow \exists s\ Emp^+(n,c,s,t)$$
$$\sigma_2^+: \forall  n,c,s,t \ E^+(n,c,t) \wedge S^+(n,s,t) \rightarrow Emp^+(n,c,s,t)$$
and the following egd:
$$\forall n, c, s, s',t \ Emp^+(n,c,s,t) \wedge Emp^+(n,c,s',t) \rightarrow s= s'$$
\end{example}

\subsection{Interval-annotated nulls}
The c-chase procedure produces a new type of unknown value for representing unknown values generated as a result of data exchange (that is, existentially quantified variables in the rhs of s-t tgds).
 The c-chase procedure cannot use labeled nulls any more. We show the insufficiency of labeled nulls with an example. Consider the concrete fact $\allowbreak Emp(Ada,IBM,N, [0, 2))$, where $N$ is a labeled null showing the salary of $Ada$ is unknown during the time interval $[0, 2)$. In the abstract view of this fact, the snapshots $db_{0}$ and $db_{1}$ contain the fact $Emp(Ada, IBM, N)$. The abstract view of this fact is shown in the Example \ref{ex:unionhomo}. In Example \ref{ex:unionhomo} we have shown that we cannot define a homomorphism from an abstract instance in which the same labeled null appears in different snapshots to an instance that has different labeled nulls in each snapshot. The chase on the abstract view generates different labeled nulls in different snapshots. In order to be able to show that the result of the chase on the concrete view has correct semantics (defined by the chase on the abstract view), we introduce {\em interval-annotated nulls}. These nulls are annotated with the time interval of the concrete facts they occur in. For example, $N^{[s,e)}$ is an interval-annotated null in a concrete fact with the time interval $[s,e)$. The concrete fact $Emp(Ada, IBM, N^{[0,2)}, [0,2))$ shows that not only the salary of $Ada$ is unknown in the time interval $[0,2)$, but also that it {\em can be different} at snapshots $db_0$ and $db_1$ (the instance $J_2$ in Figure \ref{fig:unionhomo}). As another example, consider a concrete fact $Emp(Ada, IBM, N^{[8, \infty)}, [8,\infty))$. The interval-annotated null $N^{[8,\infty)}$ represents the sequence of labeled nulls $\langle N_8, N_9 , \ldots \rangle$. In the abstract view, the snapshot $db_8$ contains the fact $Emp(Ada,\ IBM,\ N_{8})$, the snapshot $db_9$ contains the fact $\allowbreak Emp(Ada,\ IBM,\ N_{9})$ and so on. 
  
 An interval-annotated null 
is an expression $N^{[s,e)}$ where $N$ is a labeled null and
$[s,e)$ is a time interval which is the {\em temporal context of $N$}.
Each interval-annotated null $N^{[s,e)}$ (where $e \neq \infty$) corresponds to a finite sequence of distinct labeled nulls $\langle N_s$, ..., $N_{e-1} \rangle$. In case of $N^{[s,\infty)}$, the interval-annotated null corresponds to the infinite sequence $\langle N_s, N_{s+1}, ... \rangle$ of labeled nulls.  In order to choose a labeled null in the sequence of nulls represented by $N^{[s,e)}$ we project on a time point, that is $\Pi_{\ell}(N^{[s,e)}) = N_{\ell}$, $s \leq \ell < e$.  We denote by $\mathbf{N}^{[s,e)}$, a vector of interval-annotated nulls that occur in a concrete fact with the time interval of $[s,e)$. We extend $\sem{.}$ to instances with interval-annotated nulls. Let $I_c$ be a concrete instance, then $\sem{I_c}$ is a sequence of snapshots $\langle db_0, db_1, \ldots \rangle$  such that for all $\ell \in \mathbb{N}_0$:

$$db_{\ell} = \{\ R(\mathbf{a}, \Pi_{\ell}(\mathbf{N}^{[s,e)})\  |\  \exists s. \exists e.\  \ R^+(\mathbf{a}, \mathbf{N}^{[s,e)]} , [s,e)) \in I_c \text { and  } s \leq \ell < e\}$$

\subsection{Normalization}
In a concrete source instance we have the temporal attribute with time intervals as values. Chase steps use homomorphisms from the lhs of a dependency to an instance to translate data. Informally, we would like to be able to define a homomorphism from a conjunction of atomic formulas $\phi^+(\mathbf{x},t)$ to a concrete instance $I_c$ whenever there are homomorphisms from $\phi(\mathbf{x})$ to $\sem{I_c}$. 
 As an example, suppose we are trying to define a homomorphism from the lhs of $\sigma_2^+$ (in Example \ref{ex:conrunex}) to the constants and time intervals in the instance shown in Figure \ref{fig:coninstance}:
$$h: \{ n \mapsto Ada, c \mapsto IBM, s \mapsto 18k, t \mapsto ?\}$$
One cannot map the variable $t$ to a single time interval $h(t)$ such that $E^+(h(n), h(c), h(t))$ and $S^+(h(n), h(s), h(t))$ are some concrete facts in the instance $I_c$ shown in Figure \ref{fig:coninstance}. In fact no homomorphism can be defined from the lhs of $\sigma_2^+$ to $I_c$. However, if we consider the abstract view of the same data (shown in Figure \ref{fig: snapshotsrc}), many homomorphisms can be defined from $\sigma: E(n,c) \wedge S(n,s)$ to $\sem{I_c}$ including the homomorphism $h'$: 
$$h':\{n \mapsto Ada, c\mapsto IBM, s \mapsto 18k\}$$
from $\sigma$ to the snapshot of $\sem{I_c}$ associated with time point $2013$. 
\noindent \noindent We would like to have a concrete instance $I_c$ with the following property:

\begin{definition}\textbf{Normalization Property:} \label{def:normprop}
Let $I_c$ be a concrete instance and $\Phi^+$ be a set of temporal conjunctions respectively. Obtain the corresponding set of conjunctions $\Phi$ on schema of the snapshots in $\sem{I_c}$. The instance $I_c$ has the normalization property w.r.t. $\Phi^+$ when both of the following conditions hold:
\begin{itemize}
    \item \textbf{Condition 1}: $\forall \phi \in \Phi \ \forall \ell \in \mathbb{N}_0$, if $h_{\ell} : \phi(\mathbf{x}) \mapsto db_{\ell}$ ($db_{\ell} \in \sem{I_c}$), then there is a homomorphism $h$ from the conjunction of atomic formulas $\phi^+(\mathbf{x},t) \in \Phi^+$ to $I_c$ such that $\ell \in h(t)$. Also the homomorphisms $h$ and $h_{\ell}$ map the same variable $x \in \mathbf{x}$ to the same constant (that is $\forall x \in \mathbf{x}. \ h(x) = h_{\ell}(x) \text { if } h_{\ell}(x) = a, \ a \in \Const$).
    \item \textbf{Condition 2}: $\forall \phi^+ \in \Phi^+$ if $h: \phi^+(\mathbf{x},t) \mapsto I_c$ where $h(t) = [s,e)$, then there are homomorphisms $h_s, \ldots, h_{e-1}$ from $\phi(\mathbf{x})$ to consecutive snapshots $db_s, \ldots, db_{e-1}$ such that:
 \begin{itemize}
    \item $h_s:\  \phi(\mathbf{x})\mapsto db_s$,
    \item $h_{s+1}:\  \phi(\mathbf{x})\mapsto db_{s+1}$,
    \item $\ldots$,
    \item $h_{e-1}: \phi(\mathbf{x})\mapsto db_{e-1}\ $,
    \item $\forall x \in \mathbf{x} \ \text{ if } h(x) = a,~~ a \in \Const \text{, then }\ \forall j \in \{s, \ldots, e-1\} \ h_j(x)=a. $
\end{itemize}
\end{itemize}
\end{definition}

A concrete instance is {\em normalized} with respect to a set of temporal conjunctions $\Phi^+$ if it has the normalization property w.r.t. $\Phi^+$. In a normalized concrete instance the time intervals behave as constants (as shown in the Example~\ref{ex:normprop}).

\begin{example}\label{ex:normprop}
The instance $I'_c$ shown in Figure \ref{fig:normconinstance} is normalized (by fragmenting the concrete facts in $I_c$) with respect to $E^+(n,c,t) \wedge S^+(n,s,t)$ (i.e. the lhs of $\sigma_2^+$). For example, there is a homomorphism $h$ from $E^+(n,c,t) \wedge\  S^+(n,s,t)$ to the concrete instance $I_c$ such that $$h= \{n \mapsto Ada, c \mapsto Google, s \mapsto 18k, t \mapsto [2014, \infty)\}.$$ Since $I'_c$ is normalized, there are infinitely many homomorphisms $h_{\ell} , \ell \geq 2014$ from $E(n,c) \wedge S(n,s)$ to snapshots $db_\ell \in \sem{I'_c}$ such that:
 $$h_{\ell}(n) = h(n),~ h_{\ell}(c) = h(c) \text{ and } h_{\ell}(s) = h(s) ,~ \ell \in h(t)$$
Also, consider the homomorphism $h'_{\ell}$ ($\ell = 2013$) to snapshot $db_{
\ell}$:
$$h'_{\ell} = \{n \mapsto Ada, c \mapsto IBM, s \mapsto 18k\}$$ 
Since $I'_c$ is normalized, there is a homomorphism $h'$ from $E^+(n,c,t) \wedge\  S^+(n,s,t)$ to $I_c$ such that $2013 \in h'(t) = [2013, 2014)$ and 
$h'(n) = h'_{\ell}(n)$, $h'(s) = h'_{\ell}(s)$ and $h'(c) = h'_{\ell}(c)$.
\end{example}

\begin{figure}
    \centering

  \begin{tabular}{|c|c|c|}  
  \multicolumn{1}{l}{$E^+$} & \multicolumn{1}{r}{} & \multicolumn{1}{r}{}\\ \hline
  Name & Company & Time \\ \hline
   Ada & IBM &  [2012, 2013) \\ \hline
   Ada & IBM &  [2013, 2014) \\ \hline
   Ada & Google & [2014, $\infty$) \\ \hline
   Bob & IBM & [2013, 2015) \\ \hline
   Bob & IBM & [2015, 2018) \\ \hline
 
  \end{tabular}
  \quad 
  \quad
  \begin{tabular}{|c|c|c|}
    \multicolumn{1}{l}{$S^+$} & \multicolumn{1}{r}{} & \multicolumn{1}{r}{}\\ \hline
       Name&Salary& Time  \\ \hline
       Ada&18k&[2013, 2014) \\ \hline
       Ada&18k&[2014, $\infty$) \\ \hline
       Bob & 13k & [2015, 2018) \\ \hline
       Bob & 13k& [2018, $\infty$) \\ \hline
  \end{tabular}
    \caption{A normalized concrete source instance $I'_c$ w.r.t. $E^+(n,c,t) \wedge S^+(n,s,t)$}
    \label{fig:normconinstance}
\end{figure}

In the rest of this section, we discuss how to obtain a normalized instance with respect to conjunctions of atomic formulas. Note that the lhs of s-t tgds and egds (discarding the quantification) is conjunctions of atomic formulas. 

Let $\Phi^+$ be a set of temporal conjunctions of the form $\phi^+(\mathbf{x},t)$. Denote by $|\phi|$ the number of atoms that are in $\phi$. We denote by $\mathcal{N}(\Phi^+)$ the normalized form of $\Phi^+$ such that for each formula $\phi^+ \in \Phi^+$ each occurrence of the variable $t$ in $\phi^+$ is replaced with a new variable $t'$ in $\mathcal{N}(\Phi^+)$. 

Let $\Phi^+$ be temporal conjunctions of atomic formulas. Let $I_c$ be a concrete instance and $\{f_1, \ldots, f_n\} \subseteq I_c$. Let $\phi^* \in \mathcal{N}(\Phi^+)$. We denote by $h: \phi^* \mapsto \{f_1, f_2, \ldots, f_n\}$, where $|\phi^*| = n$, a homomorphism from $\phi^*$ to $I_c$ such that for every atom $R_i(\mathbf{x},t_0)$ in $\phi^*$, $R_i(h(\mathbf{x}), h(t_0))$ is $f_i$, $1 \leq i \leq n$.
\begin{example}
Let $\Phi$ contains a temporal conjunction $\phi^+= R^+(x,t) \wedge S^+(y,t)$. Then the corresponding $\mathcal{N}(\Phi)$ contains:
$$\phi^*= R^+(x,t_1) \wedge S^+(y, t_2)$$
\end{example}
The intuitive idea behind using $\mathcal{N}(\Phi^+)$ instead of $\Phi^+$ is to be able to map the temporal variable in each atom in a conjunction in $\mathcal{N}(\Phi^+)$ to a different time interval. 

\begin{definition}\textbf{Empty intersection property}\label{def:emptyinterprop}
A concrete instance $I_c$ has the {\em empty intersection property} with respect to a set of temporal conjunctions $\Phi^+$ if for every homomorphism $h$ from a conjunction of atomic formulas $\phi \in \mathcal{N}(\Phi^+)$ to $I_c$ such that $h: \phi^* \mapsto \{f_1, f_2, \ldots, f_n\}$, then
\begin{enumerate}
    \item either $(\bigcap_{i \in \{1, \ldots, n\}} f_i[T]) = \emptyset $
    \item or, $\bigcap_{i \in \{1, \ldots, n\}} f_i[T] = \bigcup_{i \in \{1, \ldots, n\}} f_i[T]$
\end{enumerate}
\end{definition}

In the next theorem we will show that an instance has the normalization property with respect to conjunctions of atomic formulas if and only if it has the empty intersection property. 
\begin{theorem}\label{thm:normprop}
Let $\Phi^+$ be a set of temporal conjunctions. A concrete instance $I_c$ is normalized with respect to $\Phi^+$ if and only if $I_c$ has the empty intersection property with respect to $\Phi^+$.
\end{theorem}
\begin{proof}
\noindent\textbf{{\em The if direction.}} 
In this direction, the concrete instance $I_c$ is normalized and we show $I_c$ has the empty intersection property as well. 
Let $h$ be a homomorphism from $\phi^+(\mathbf{x},t) \in \Phi^+$ to the instance $I_c$ such that the images of the atoms in $\phi^+$ under $h$ are the concrete facts $f_1, ..., f_n$, where $n = |\phi^+|$. The temporal variable $t$ (under $h$) has to map to a single interval $h(t) = [s,e)$ (otherwise a homomorphism cannot be defined). This means that $\forall i \in \{1, \ldots, n\}, f_i[T] = [s,e)$. Let $\phi^* \in \mathcal{N}(\Phi^+)$ be a conjunction of atomic formulas that is obtained by replacing each occurrence of the temporal variable $t$ in $\phi^+(\mathbf{x},t)$ with a new variable. Define $h'$ as follows:
$$h'(x) = h(x), \forall x \in \mathbf{x} \text{ and } h'(t') = h(t), t' \text{ a temporal variable in } \phi^*$$

We have $h': \phi^* \mapsto \{f_1, \ldots, f_n\}$ (because $h:\phi^+ \mapsto \{f_1, \ldots, f_n\}$ and by construction of $\phi^*$). Since $\forall i \in \{1, \ldots, n\}, f_i[T] = [s,e)$,  we have $\bigcap_{i \in \{1, \ldots, n\}} f_i[T] = \bigcup_{i \in \{1, \ldots, n\}} f_i[T]$. Thus, $I_c$ has the empty intersection property.

\noindent\textbf{{\em The only if direction}}
Consider a $\phi^* \in \mathcal{N}(\Phi^+)$. Let $\phi^+$ be the corresponding temporal conjunction of atomic formulas with the same temporal variable in each atom. Let $B = \{f_{c_1}, \ldots, f_{c_n}\}$ be a subset of $I_c$. Let $h'$ be a homomorphism $h': \phi^* \mapsto B$. Since $I_c$ has the empty intersection property the time interval of the facts in $B$ are either equal or the intersection of the time intervals is empty. In the latter case, no homomorphism can be defined from $\phi^+(\mathbf{x},t)$ to $I_c$  because the variable $t$ in each atom cannot be mapped to a single interval. Therefore, we just consider the former case (that is the $\bigcup_{f_c\in B} f_c[T] = \bigcap_{f_c \in B} f_c[T]$). This means all the facts in $B$ have the same time interval, that is $\forall f_c \in B \ f_c[T] = [s,e)$. Consider any concrete fact $f_c \in B$. W.l.o.g. we assume the interval-annotated nulls in the fact $f_c$ are preceded by all the constants, that is $$f_c = R(\mathbf{a}, \mathbf{N}^{[s,e)}, [s,e)).$$ By definition of $\sem{.}$, for each $f_{c_i} \in B$, $1 \leq i \leq n$, the snapshot $db_{\ell} \in \sem{I_c}$ ($s \leq \ell < e$), contains the fact $$f_{a_i}: R(\mathbf{a}, \Pi_{\ell}(\mathbf{N^{[s,e)}})).$$  Let $\phi$ be the corresponding conjunction of $\phi^+$ over the snapshots. We need to show $I_c$ has the normalization property. 
Define $h$ as follows:
$$h(x) = h'(x), \forall x \in \mathbf{x} \text{ and } h(t) = [s,e)$$
Since $h'$ is a homomorphism from $\phi^*$ to $B$ and all the concrete facts in $B$ has the same time interval, $h$ is a homomorphism from $\phi^+$ to $I_c$ such that the image of each atom $R^+_i(\mathbf{x},t)$ under $h$ is a fact in $B$. Define homomorphisms $h_s , \ldots, h_{e-1}$ from $\phi(\mathbf{x})$ to consecutive snapshots $db_s$ to $db_{e-1}$.
For each $\ell \in \{s, \ldots, e-1\}$, define
\[h_{\ell}(x) = 
\left\{
	\begin{array}{ll}
	h(x) & \mbox{if } h(x) \mbox{ is a constant } \\ 
	\Pi_{\ell}(N^{[s,e)}) & \mbox{if } h(x) \mbox{ is an interval-annotated null }  N^{[s,e)}\\
	\end{array}
\right.\]
 For each atom $R_i(\mathbf{x})$ in $\phi$, the image of the atom under $h_{\ell}$, that is $R_i(h_{\ell}(\mathbf{x})) = R(\mathbf{a}, \mathbf{N}_{\ell})$ which is the fact $f_{a_i}$ in the snapshot $db_{\ell}$ ($s \leq \ell < e$). Thus, condition 2 of the normalization property holds. 
 
Consider any homomorphism $h_{\ell}$, $s \leq \ell < e$. by definition of $h'$ and $h_{\ell}$ it follows that:
$$\forall x \in \mathbf{x} \text{ if } h_{\ell}(x) =a \text{ then } h_{\ell} (x) = h(x)$$
Also, $\ell \in h(t) = [s,e)$. Thus, condition 1 in the definition of the normalization property holds as well.
 Therefore, $I_c$ is normalized. 
\end{proof} 

Let $\Phi^+$ be a set of temporal conjunctions. Let $I_c$ be a concrete instance with $n$ facts that is not normalized w.r.t. $\Phi^+$. We show in Theorem \ref{thm:normsize} that the size of a normalized instance w.r.t. $\Phi^+$ (obtained by fragmenting the concrete facts in $I_c$) is $\mathcal{O}(n^2)$. Example \ref{ex:normintuition} discusses intuitions behind Theorem \ref{thm:normsize}. 
\begin{example}\label{ex:normintuition}
Suppose $I_c$ is a concrete instance with two facts $f_1$ and $f_2$. Let $\Phi^+$ contains a temporal conjunction of atomic formulas $\phi^+$ (over schema of $I_c$). Suppose $I_c$ is not normalized and there is a homomorphism from $\phi^*$ to $\{f_1, f_2\}$ where $\phi^* \in \mathcal{N}(\Phi^+)$ is the corresponding conjunction for $\phi^+$ (with different temporal variables in each atom). Let $f_1[T] = [s_1, e_1)$ and $f_2[T] = [s_2, e_2)$. Since $I_c$ is not normalized $f_1[T] \cap f_2[T] \neq \emptyset$. Since the time intervals overlap, one of the following cases holds:
\begin{itemize}
\item $s_1<s_2<e_1<e_2$
\item $s_2 < s_1 < e_2 < e_1$
\item $s_1 < s_2 < e_2 < e_1$
\item $s_2 < s_1 < e_1 < e_2$
\end{itemize}
 Here we consider that $f_1[T]$ overlaps with $f_2[T]$ according to the first case. We fragment the facts $f_1$ and $f_2$ so that the normalized instance satisfy the empty intersection property: 
\begin{itemize}
    \item $f_{11}$, where $f_{11}[T] = [s_1, s_2)$.
    \item $f_{12}$, where $f_{12}[T] = [s_2, e_1)$.
    \item $f_{21}$, where $f_{21}[T] = [s_2, e_1)$.
    \item $f_{22}$, where $f_{22}[T] = [e_1, e_2)$.
\end{itemize}
The data attribute values of $f_{11}$ and $f_{12}$ (resp. $f_{21}$ and $f_{22}$) are same as $f_1$ (resp. $f_2$). Observe that any pair of the fragmented facts above satisfies $\phi^*$ and has disjoint or equal time intervals. The other cases of overlap can be resolved in a similar way.
\end{example}
We assume whenever we fragment a concrete fact, the annotation of an interval-annotated null in the concrete fact is changed in the fragmented facts such that the annotation is always equal to the time interval of the fact the interval-annotated null occurs in. So if $f_1$ contains an interval-annotated null $N^{[s_1,e_1)}$, then $f_{11}$ and $f_{12}$ contain interval-annotated nulls $N^{[s_1, s_2)}$ and $N^{[s_2, e_1)}$ respectively. Each of the facts $f_1$ and $f_2$ is fragmented into two facts (with smaller time intervals).
\begin{theorem}\label{thm:normsize}
Let $I_c$ be a concrete instance with $n$ facts that is not normalized w.r.t. a set of temporal conjunctions $\Phi^+$. Let $I'_c$ be a concrete instance that is obtained by fragmenting the facts in $I_c$ such that $I'_c$ is normalized w.r.t. $\Phi^+$. The size of $I'_c$ is $\mathcal{O}(n^2)$ if each fact in $I_c$ needs to be fragmented. 
\end{theorem}
\begin{proof}
In general, a fact $f \in I_c$ (that needs to be fragmented) with time interval $[s_i, e_i)$ is fragmented into $k_i$ number of facts such that $k_i$ is the number of distinct start points and endpoints that are greater than or equal to $s_i$ and smaller than $e_i$:
\[
\underbrace{s_i<...<s_j<...<e_m<...<}_{k_i}e_i
\]
An instance $I_c$ that contains $n$ concrete facts have at most $2n$ distinct start points and end points. Therefore, $k_i \leq 2n-1$. In the worst case (which depends on the set of conjunctions and the time interval of the facts that satisfy a conjunction of atomic formulas) each concrete fact needs to be fragmented considering all the distinct start points and end points in the instance. Therefore, the normalized instance is of size $\mathcal{O}(n^2)$.
\end{proof}
A na\"{i}ve normalization algorithm, fragments each fact without considering any conjunction of atomic formulas (as if $\Phi^+ = \emptyset$) and only based on the start points and end points of all the other facts. Such an algorithm needs to sort the start points and endpoints of all the facts. Thus, the time complexity of a na\"{i}ve normalization algorithm is $\mathcal{O}(nlogn)$ where $n$ is the number of facts in the original (non-normalized) instance. However, a na\"{i}ve normalization algorithm generates unnecessary fragments when there is no homomorphism from a conjunction of atomic formulas to any subset of the facts. Figure \ref{fig:naivenorm} depicts a normalized instance w.r.t. $\Phi^+ = \emptyset$ generated by a nai\"ve normalization algorithm. Observe that the normalized instance w.r.t. $E^+(n,c,t) \wedge S^+(n,s,t)$ shown in Figure \ref{fig:naivenorm} has more facts compared to the normalized instance shown in Figure \ref{fig:normconinstance}. The reason is that the normalized instance in Figure \ref{fig:normconinstance} is obtained by taking the conjunction $E^+(n,c,t) \wedge S^+(n,s,t)$ into consideration during fragmentation of facts. In the remaining of this sub-section we discuss our proposed algorithm for normalizing an instance w.r.t. temporal conjunctions. 

\begin{figure}
    \centering

  \begin{tabular}{|c|c|c|}  
  \multicolumn{1}{l}{$E^+$} & \multicolumn{1}{r}{} & \multicolumn{1}{r}{}\\ \hline
  Name & Company & Time \\ \hline
   Ada & IBM &  [2012, 2013) \\ \hline
   Ada & IBM &  [2013, 2014) \\ \hline
   Ada & Google & [2014, 2015) \\ \hline
   Ada & Google & [2015, 2018) \\ \hline
   Ada & Google & [2018, $\infty$) \\ \hline
   Bob & IBM & [2013, 2014) \\ \hline
   Bob & IBM & [2014, 2015) \\ \hline
   Bob & IBM & [2015, 2018) \\ \hline
  \end{tabular}
  \quad 
  \quad
  \begin{tabular}{|c|c|c|}
    \multicolumn{1}{l}{$S^+$} & \multicolumn{1}{r}{} & \multicolumn{1}{r}{}\\ \hline
       Name&Salary& Time  \\ \hline
       Ada&18k&[2013, 2014) \\ \hline
       Ada&18k&[2014, 2015) \\ \hline
       Ada&18k&[2015, 2018) \\ \hline
       Ada&18k&[2018, $\infty$) \\ \hline
       Bob & 13k & [2015, 2018) \\ \hline
       Bob & 13k& [2018, $\infty$) \\ \hline
       
  \end{tabular}
    \caption{A normalized concrete source instance obtained by a nai\"ve normalization algorithm}
    \label{fig:naivenorm}
\end{figure}

We propose an algorithm that fragments the concrete facts in an instance based on $\Phi^+$. The normalization algorithm $norm(I_c, \Phi^+)$ (Algorithm \ref{alg:normalization}) receives a concrete instance $I_c$ and $\mathcal{N}(\Phi^+)$, fragments the concrete facts in $I_c$ and returns a normalized concrete instance $I'_c$ w.r.t. $\Phi^+$.
The algorithm first builds the set $\mathcal{S}$ which is a set of sets of concrete facts in $I_c$ that satisfy some formula $\phi^+ \in \mathcal{N}(\Phi^+)$. Then the sets that have at least a concrete fact in common are moved to another set $S_{\cap}$. The sets that are in $S_{\cap}$ and have a concrete fact in common are merged until no more merges can be done. After adding the merged sets to $\mathcal{S}$, the concrete facts that are in each set $\Delta \in \mathcal{S}$ are fragmented by sorting the time intervals of the concrete facts in $\Delta$ and fragmenting the time intervals such that they do not overlap anymore. Example \ref{ex:normalizaion} shows how algorithm $norm(I_c, \Phi^+)$ works.
\begin{figure}
    \centering
    \begin{tabular}{|c|c|c|}
     \multicolumn{1}{l}{$R^+$} & \multicolumn{1}{r}{} & \multicolumn{1}{r}{}\\ \hline
         & A & T \\ \hline
        $f_1$ & a & [5,11) \\ \hline
    \end{tabular}
    \quad 
    \quad
        \begin{tabular}{|c|c|c|}
     \multicolumn{1}{l}{$P^+$} & \multicolumn{1}{r}{} & \multicolumn{1}{r}{}\\ \hline
         & A & T \\ \hline
        $f_2$ & a & [8,15) \\ \hline
        $f_4$ & b & [20, 25) \\ \hline
    \end{tabular}
    \quad 
    \quad
     \begin{tabular}{|c|c|c|}
     \multicolumn{1}{l}{$S^+$} & \multicolumn{1}{r}{} & \multicolumn{1}{r}{}\\ \hline
         & A & T \\ \hline
        $f_3$ & a & [7,10) \\ \hline
        $f_5$ & b & [18,$\infty$) \\ \hline
    \end{tabular}
    \caption{Input of the normalization algorithm in Example \ref{ex:normalizaion}}
    \label{fig:normalgex}
\end{figure}

\SetAlgoLined
\begin{algorithm}
\small
\SetKwInOut{Input}{Input}\SetKwInOut{Output}{Output}
\Input{Concrete instance $I_c$ and $\Phi^+$.}
\Output{Normalized instance $I'_c$ w.r.t. $\Phi^+$}
\BlankLine
       $I'_c = I_c$\; 
       build $\mathcal{N}(\Phi^+)$ \; 
           $\begin{aligned}
                & \mathcal{S} = \{\Delta \ | \  \Delta = \{f_1, \ldots, f_m\}, f_1, \ldots, f_m \in I_c \text{ such that }  \bigcap_{f \in \Delta} f[T] \neq \emptyset & \\ & \text{ and } \exists \phi^+ \text{ such that } \phi^+ \in \mathcal{N}(\Phi^+) \\ & \text{ and }
            \  m = |\phi^+|\   
             \text{ and there is a homomorphism h}  \text{ s.t. } h:\phi^* \mapsto \Delta \}
           \end{aligned}$\;
       $S_{\cap} = \{\Delta \in \mathcal{S}\  |\  \exists \Delta' \in \mathcal{S}. \exists f \ \text{ such that } f \in (\Delta \cap \Delta')\}$\;
       \vspace{0.25cm}
       $\mathcal{S} = \mathcal{S} \backslash S_{\cap}$\;
       \vspace{0.20cm}
       \While{$\exists \Delta_1, \Delta_2 \in S_{\cap} \text{ such that } (\Delta_1 \neq \Delta_2 \text{ and } \Delta_1 \cap \Delta_2 \neq \emptyset)$}{
       $\Delta' = \Delta_1 \cup \Delta_2$\;
       $S_{\cap} = (S_{\cap} \backslash \Delta_1, \Delta_2) \cup \{\Delta'\}$\;}
       \vspace{0.10cm}
       $\mathcal{S} = \mathcal{S} \cup S_{\cap}$\;
       \vspace{0.10cm}
       \For{each $\Delta \in \mathcal{S}$}{
       $TP_{\Delta} =\langle tp_1, tp_2,  \ldots, tp_m \rangle$, where $tp_i$ is a distinct start point or end point in the facts in $\Delta$ and $m$ is the number of distinct start points and end points in $\Delta$\;
       Sort $TP_{\Delta}$ in ascending order of time points\;
       \For{each $f \in \Delta$ such that $f[T] = [s_i, e_i)$:}
       {$TP_{f} = \langle s_i, \ldots,  e_i \rangle$ is a sub-sequence of $TP_{\Delta}$ from the time point $s_i$ to the time point $e_i$\;
       $k = |TP_{f}| - 1$\;
       Fragment the fact $f_c$ to $k$ facts such that
       $$frg = \forall j\in \{1,\ldots, k\}~\  f_j[\mathbf{D}] = f_c[\mathbf{D}] \text { and } f_j[T] = [TP_f[j],\  TP_f[j+1])\;$$
       \vspace{0.1cm}
       $I'_c = I'_c \backslash \{f\} \cup \ frg \;$

       }
     }    
       \Output{$I'_c$}
\caption{\label{alg:normalization}$norm(I_c,\Phi^+)$}
\end{algorithm}

\begin{example}\label{ex:normalizaion}
Consider a schema with three relation symbols $R^+$, $P^+$, $S^+$ each with the attributes $A$ and $T$. Consider an instance of $I_c$ of this schema with five facts as shown in Figure \ref{fig:normalgex}. 
Let $\mathcal{N}(\Phi^+)$ contains two conjunctions of atomic formulas: 
$$ \phi_1: R^+(x,t_1) \wedge P^+(y, t_2) 
\text{ and }$$
$$\phi_2: P^+(x, t_1) \wedge S^+(y, t_2)$$

The algorithm first builds the set $\mathcal{S}$. In this example $R$ is the only relation in the instance, so
$$\mathcal{S} = \{\{f_1, f_2\}, \{f_2, f_3\}, \{f_4, f_5\}\}$$
Each set $\Delta$ in $\mathcal{S}$ satisfies a conjunction of atomic formulas (treating time intervals as constants) and in each $\Delta$ the intersection of the time intervals of the facts is not empty.
The algorithm continues by building the set $S_{\cap}$ which is a subset of $\mathcal{S}$ and contains the sets of facts that have a common fact with each other. In this example $$S_{\cap} = \{\{f_1, f_2\}, \{f_2, f_3\}\}.$$  
After building $S_{\cap}$, the algorithm removes the sets in $S_{\cap}$ from $\mathcal{S}$ and merges the sets in $S_{\cap}$ that have common facts. In this example after merging the sets in $S_{\cap}$ we have:
$$S_{\cap} = \{\{f_1,f_2, f_3\}\}$$
After adding $S_{\cap}$ to $\mathcal{S}$:
$$\mathcal{S} = \{\{f_1,f_2, f_3\}, \{f_4, f_5\}\}$$
In this example there are two sets $\Delta_1$ and $\Delta_2$ in $\mathcal{S}$. 
The algorithm sorts the distinct start points and end points of the facts in $\Delta_1$ and $\Delta_2$:
\begin{itemize}
    \item $TP_{\Delta_1} : \langle 5,7, 8, 10, 11,  15 \rangle$
    \item $TP_{\Delta_2} : \langle  18, 20 , 25, \infty \rangle$
\end{itemize}
Here we just show how the fact $f_1: R^+(a,[5,11))$ is fragmented:
\begin{itemize}
    \item $f_{11}$, where $f_{11}[T] = [5, 7)$
    \item $f_{12}$, where $f_{12}[T] = [7, 8)$
    \item $f_{13}$, where $f_{13}[T] = [8, 10)$
    \item $f_{14}$, where $f_{14}[T] = [10, 11)$
\end{itemize}
At the end the algorithm removes $f_1$ from the instance $I'_c$ and adds the fragmented facts. The other facts in $\Delta_1$ and $\Delta_2$ are fragmented the same way as well. The final normalized instance is shown in Figure \ref{fig:outputnormalized}.
\end{example}
\begin{figure}
    \centering
    \begin{tabular}{|c|c|c|}
     \multicolumn{1}{l}{$R^+$} & \multicolumn{1}{r}{} & \multicolumn{1}{r}{}\\ \hline
         & A & T \\ \hline
        $f_{11}$ & a & [5,7) \\ \hline
        $f_{12}$ & a& [7,8) \\ \hline
        $f_{13}$ & a &[8,10) \\ \hline
        $f_{14}$ & a & [10,11) \\ \hline
    \end{tabular}
    \quad 
    \quad
        \begin{tabular}{|c|c|c|}
     \multicolumn{1}{l}{$P^+$} & \multicolumn{1}{r}{} & \multicolumn{1}{r}{}\\ \hline
         & A & T \\ \hline
        $f_{21}$ & a & [8,10) \\ \hline
        $f_{22}$ & a & [10,11) \\ \hline
        $f_{23}$ & a & [11,15) \\ \hline
        $f_4$ & b & [20, 25) \\ \hline
    \end{tabular}
    \quad 
    \quad
     \begin{tabular}{|c|c|c|}
     \multicolumn{1}{l}{$S^+$} & \multicolumn{1}{r}{} & \multicolumn{1}{r}{}\\ \hline
         & A & T \\ \hline
        $f_{31}$ & a & [7,8) \\ \hline
         $f_{31}$ & a & [8,10) \\ \hline
        $f_{51}$ & b & [18,20) \\ \hline
        $f_{52}$ & b & [20,25) \\ \hline
        $f_{53}$ & b & [25,$\infty$) \\ \hline
    \end{tabular}
    \caption{Output of the normalization algorithm}
    \label{fig:outputnormalized}
\end{figure}

\begin{theorem}
Let $I'_c = norm(I_c, \Phi^+)$. The instance $I'_c$ is normalized. 
\end{theorem}
\begin{proof}
We will show $I'_c$ has the empty intersection property. Therefore, based on Theorem \ref{thm:normprop}, $I'_c$ is normalized.

Let $\phi^*$ be a conjunction of atomic formulas in $\mathcal{N}(\Phi^+)$. Let $h$ be a homomorphism from $\phi^*$ to a set of concrete facts $f_{c_1}, \ldots, f_{c_n}$ in $I'_c$. We need to show either $(\bigcap_{i \in \{1,..,n\}} f_{c_i}[T]) = \emptyset$ or $\bigcap_{i \in \{1,..,n\}} f_{c_i}[T] = \bigcup_{i \in \{1,..,n\}} f_i[T]$.

Let $B = \{f_{c_1}, \ldots f_{c_n}\}$. 
Denote by $br(f)$ the set of fragmented facts of a concrete fact $f \in I_c$ obtained by the algorithm. Each fact $f_{c_i}$ in $B$ is either obtained by fragmenting a concrete fact $f_i$ in $I_c$ (that is $f_{c_i} \in br(f_i)$) or is a concrete fact in $I_c$ (that is $f_{c_i} = f_i$).
Define $h'$ to be $h$ except that the temporal variable $t_i$ in each atom $R_i$ is mapped to $h'(t_i) = f_i[T]$. Since $f_{c_i}[\mathbf{D}] = f_i[\mathbf{D}]$ and the temporal attribute values of $f_{c_i}$ and $f_i$ do not matter when considering $\phi^* \in \mathcal{N}(\Phi^+)$, $h'$ is a homomorphism from $\phi^*$ to $\{f_1, f_2, \ldots f_n\}$.

If $(\bigcap_{i \in \{1,..,n\}} f_i) = \emptyset$, then it is obvious that $(\bigcap_{i \in \{1,..,n\}} f_{c_i}[T]) = \emptyset$. So we consider the case that 
$(\bigcap_{i \in \{1,..,n\}} f_i) \neq \emptyset$. Thus, there is a set in $\mathcal{S}$ such that $f_1, \ldots f_n \in \Delta$.

Let $TP_{\Delta}$ contain the sorted distinct start points and end points of the facts in $\Delta$, that is $\langle \ell_1, \ldots, \ell_m \rangle$, where $m$ is the number of distinct start points and end points in $\Delta$. Consider an arbitrary fact $f_{c_i}$ in $B$. Let $f_{c_i}[T] = [s, e)$. By construction of the fact $f_{c_i}$ by the algorithm (which is obtained by fragmenting a concrete fact $f_i \in \Delta$), we have $\langle \ell_1, \ldots, s, e , \ldots, \ell_m \rangle$. Observe that time point $e$ is the immediate timepoint after $s$ in $TP_{\Delta}$. Therefore, if any fact $f'$ in $B$ has another start point (that is $f'[T] = [s',e')$ and $s \neq s'$), then the $\bigcap_{f\in \Delta} f[T] = \emptyset$. If all the facts in $B$ have the same interval then $\bigcup_{f \in B} f[T] = \bigcap_{f \in B} f[T]$. Therefore, $I'_c$ has the empty intersection property. 

\end{proof}

The time complexity of the normalization algorithm $norm(I_c, \Phi^+)$ by the assumption of fixing $\Phi^+$ is polynomial in the size of $I_c$. Na\"ive normalization algorithm has a better time complexity ($\mathcal{O}(nlogn)$) but the size of the normalized instance is possibly larger because of the possibility of unnecessary fragments caused by not considering the schema mapping. In general there is a trade off between the cardinality of a normalized instance and the time complexity of a normalization algorithm. A more complete characterization of the applicability of the algorithms would be a subject of future work.

\subsection{Concrete chase}
Putting everything together, in this section we define the concrete chase. Considering the lhs of all s-t tgds, first the concrete source instance needs to be normalized w.r.t. the lhs of the s-t tgds. Then all {\em s-t tgd c-chase steps} are applied sequentially to get a target instance. Then the target instance needs to be normalized w.r.t. the lhs of the egds. Finally a {\em concrete solution} is obtained by applying a successful sequence of {\em egd c-chase steps}. In the rest of this chapter, whenever we say a concrete instance is normalized w.r.t. $\Sigma_{st}$ (resp. $\Sigma_{eg}$) it means it is normalized w.r.t. the lhs of $\Sigma_{st}$ (resp. $\Sigma_{eg}$). The lhs of the s-t tgds and egds are considered as conjunctions of atomic formulas.   
\begin{definition}{\em c-chase step:}
\begin{itemize}
    \item {\em (s-t tgd):} Let $\sigma^+ : \forall \mathbf{x}, t \ \phi^+(\mathbf{x}, t) \rightarrow \exists \mathbf{y} \ \psi^+ (\mathbf{x}, \mathbf{y}, t)$ be an s-t tgd. Let $I_c$ be the concrete normalized source instance and  $J_c$ be a concrete target instance (initially $J_c = \emptyset$). Let $h$ be a homomorphism from lhs of $\sigma$ to $I_c$ such that there is no extension $h'$ of $h$ from $\phi^+(\mathbf{x}, t) \wedge \psi^+ (\mathbf{x}, \mathbf{y}, t)$ to $(I_c, J_c)$. We say $\sigma^+$ {\em can be applied to $(I_c, J_c)$ with $h$}. Let $J'_c$ be the union of $J_c$ with the set of facts obtained by (a) extending $h$ to $h'$ so that each variable in $\mathbf{y}$ is assigned to a fresh null annotated with $h(t)$, followed by (b) applying $h'$ to the rhs of $\sigma^+$. We say the result of {\em applying $\sigma^+$ to $(I_c, J_c)$ is $(I_c, J'_c)$} and write $(I_c, J_c) \xmapsto{\sigma^+,h} (I_c, J'_c)$.
    \item {\em (egd):} Let $\sigma^+: \forall \mathbf{x}, t\  \phi^+(\mathbf{x}, t) \rightarrow x_1 = x_2 $ be an egd. Let $J_c$ be the concrete normalized target instance such that $(I_c, J_c) \models \Sigma^+_{st}$. Let $h$ be a homomorphism from $\sigma$ to $J_c$ such that $h(x_1) \neq h(x_2)$. We say that $\sigma^+$ {\em can be applied to $J_c$ with $h$}. We distinguish two cases:
    \begin{itemize}
        \item If both $h(x_1)$ and $h(x_2)$ are constants then {\em the result of applying $\sigma^+$ to $J_c$ with $h$} is a {\em failure} and it is denoted by $J_c \xmapsto{\sigma^+,h} \bot$.
        \item Otherwise, let $J'_c$ be $J_c$ where we identify $h(x_1)$ and $h(x_2)$ as follows: if one is a constant, then the interval-annotated null is replaced everywhere by the constant; if both are
interval-annotated nulls, then one is replaced everywhere by the other. We say $J'_c$ is {\em the result of applying $\sigma^+$ to $J_c$ with $h$}, denoted by $J_c \xmapsto{\sigma^+,h} J'_c$.
    \end{itemize}
\end{itemize}
\end{definition}
Note that in an egd c-chase step, the annotated nulls have the same time interval because the only way a homomorphism can be defined to the lhs of an egd chase step is to map variable $t$ to a single time interval. We assumed that all the interval annotated nulls in a fact are annotated with the fact's time interval.

A concrete chase is a finite sequence of s-t tgd chase steps followed by egd chase steps. We call the result of a successful concrete chase a {\em concrete solution}. If an egd c-chase step fails, then the result of c-chase is a failure.

\begin{example}
The result of concrete chase on the concrete input instance shown in Figure \ref{fig:coninstance} with the schema mappings in the Example \ref{ex:conrunex} is shown in Figure \ref{fig:conchaseresult}.
\begin{figure}
    \centering

  \begin{tabular}{|c|c|c|c|}  
  \multicolumn{1}{l}{$EMP^+$} & \multicolumn{1}{r}{} & \multicolumn{1}{r}{} & \multicolumn{1}{r}{}\\ \hline
  Name & Company & Salary & Time \\ \hline
   Ada & IBM &  $N^{[2012, 2013)}$ & [2012, 2013) \\ \hline
   Ada & IBM & 18k & [2013, 2014) \\ \hline
   Ada & Google & 18k & [2014, $\infty$) \\ \hline
   Bob & IBM & $M^{[2013, 2015)}$ &[2013, 2015) \\ \hline
   Bob & IBM & 13k &[2015, 2018) \\ \hline
  \end{tabular}
    \caption{The concrete view of c-chase($I_c, \mathcal{M}^+$)}
    \label{fig:conchaseresult}
\end{figure}
\end{example}

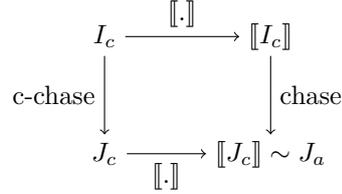
\begin{figure}

\centering

  \begin{tikzpicture}
    \matrix (m) [
      matrix of math nodes,
      row sep=3em,
      column sep=3em,
      text height=1.5ex,
      text depth=0.25ex
    ] {
      I_c & \sem{I_c} \\
           J_c & \sem{J_c} \sim J_a  \\
    };
    \path[->]        (m-1-1) edge node [above] {$\llbracket.\rrbracket$}            (m-1-2)
                     (m-1-1) edge node [left] {c-chase} 					(m-2-1)
                     (m-1-2) edge node [right] {chase }                (m-2-2)
                     (m-2-1) edge  node [below]
                     {{$\llbracket.\rrbracket$} }                 (m-2-2);
   
  \end{tikzpicture}
  \caption{ Correspondence between concrete chase on $I_c$ and chase on $\sem{I_c}$ }
   \label{fig:comsttgd}
\end{figure}

We have shown the result of a successful chase on the abstract view is a universal solution (Proposition \ref{prop:uniabstract}). Since in practice concrete instances are used the aim is to show the result of c-chase on a concrete instance has the correct semantics as if we were able to apply chase on the abstract view of that instance. This is shown in Figure \ref{fig:comsttgd}. Following the Fagin et al. \cite{FKMP05} approach to  prove that the result of the chase procedure is a universal solution, we first show a property of a c-chase step in Lemma \ref{lem: cchasestep}.  Using this lemma we show in Theorem \ref{thm:conunisol} that if $J_c$ is the result of a successful c-chase, then $\sem{J_c}$ is a universal abstract solution. 
The proof steps in Lemma \ref{lem: cchasestep} follows the proof steps of Lemma 3.4 in \cite{FKMP05}. The main difference is that the notion of homomorphism is defined from abstract instances to abstract instances (not on concrete instances). Meanwhile, a concrete chase uses homomorphisms from a temporal dependency to a concrete instance. Therefore, in the proof we have to deal with a homomorphism from a temporal dependency to an instance and its effects on a homomorphism from an abstract instance to another abstract instance.

\begin{lemma}\label{lem: cchasestep}
Let $K_c=(I_c, J_c)$ be a concrete normalized instance w.r.t. a dependency $\sigma^+$. Let $K_c \xmapsto{\sigma^+, h} K'_c$ be a chase step. Let $K_a = \langle db''_0, db''_1, \ldots \rangle$ be an abstract instance such that $K_a \models \sigma$ and $h' : \sem{K_c} \mapsto K_a$. Then there is a homomorphism $g: \sem{K'_c} \mapsto K_a$.
\end{lemma}
\begin{proof}
Let $\sem{K_c} = \langle db_0, db_1, \ldots \rangle$.
Let $\sem{K'_c} = \langle db'_0, db'_1, \ldots \rangle$. Having the homomorphism $h': \sem{K_c} \mapsto K_a$ means that there is a homomorphism from each snapshot in $\sem{K_c}$ to the corresponding snapshot in $K_a$, $$h'_{\ell}: db_{\ell} \mapsto db''_{\ell}.$$ 

{\em Case 1}: {\em $\sigma^+$ is an s-t tgd}. Then 
$h: \phi^+(\mathbf{x}, t) \mapsto K_c$.
Suppose $h(t) = [s,e)$. Based on Theorem \ref{thm:normprop} there are homomorphisms $h_s, \ldots, h_{e-1}$ from $\sigma= \phi(\mathbf{x}) \rightarrow \exists \mathbf{y} \psi(\mathbf{x}, \mathbf{y})$ to $db_s, \ldots db_{e-1}$, respectively in $\sem{K_c}$. Consider $h_{\ell}$ ($s \leq \ell < e$):
$$h_{\ell}: \phi(\mathbf{x}) \mapsto db_{\ell}.$$ 
Composing homomorphisms yields homomorphisms, thus:
$$h'_{\ell} \circ h_{\ell}: \phi(\mathbf{x}) \mapsto db''_{\ell}.$$
Since $db''_{\ell} \models \sigma$, then there exists a homomorphism $h''$ such that 
$$h''_{\ell}: \phi(\mathbf{x}) \wedge \psi(\mathbf{x}, \mathbf{y}) \mapsto db''_{\ell},$$
where $h''_{\ell}$ is an extension of $h'_{\ell} \circ h_{\ell}$ such that $h''_{\ell}(\mathbf{x}) = h'_{\ell}(h_{\ell}(\mathbf{x})).$ For each variable $y \in \mathbf{y}$ in $\psi^+$, a fresh interval-annotated null is generated in $K'_c$ that is annotated with $h(t) = [s,e)$. Denote by $N'^{[s,e)}$ the interval-annotated null generated in the chase step $K_c \xmapsto{\sigma^+, h} K'_c$. Therefore, by definition of $\sem{.}$, there is a labeled null $N'_{\ell}$ in $db'_{\ell} \in \sem{K'_c}$ ($s \leq \ell <e$).
Define $g_{\ell}$ on $\Null(db'_{\ell})$ as follows:
$g_{\ell}(N_{\ell}) = h'_{\ell}(N_{\ell})$, if $N_{\ell} \in \Null(db_{\ell})$, and $g_{\ell}(N'_{\ell}) = h''_{\ell}(y)$ for $y \in \mathbf{y}$.

In order to show $g: \sem{K'_c} \mapsto K_a$ we need to show there is a homomorphism between the corresponding snapshots. Hence, we need to show that $g_{\ell}$ is a homomorphism from $db'_{\ell}$ to $db''_{\ell}$ , $s \leq \ell < e$. 
For the facts of $db'_{\ell}$ that are also in $db_{\ell}$ this is true because $h'$ is a homomorphism; thus $h'_{\ell}: db_{\ell} \mapsto db''_{\ell}$.
Let $R^+(\mathbf{x_0}, \mathbf{y_0}, t)$ be an arbitrary atom in $\psi^+$. Therefore, the atom $R(\mathbf{x_0}, \mathbf{y_0})$ is in $\psi$.  Then $R^+(h(\mathbf{x_0}), h(\mathbf{y_0}), h(t))$ is a fact in $K'_c$. By definition of $\sem{.}$ there is a fact $R(h(\mathbf{x_0}), \pi_{\ell}(h(\mathbf{y_0})) = R(h(\mathbf{x_0}), \mathbf{N_{y_{\ell}}})$ in $db'_{\ell}$. Based on Theorem \ref{thm:normprop} we showed $ h(\mathbf{x_0}) = h_{\ell}(\mathbf{x_0}) $. By replacing $h$ with $h_{\ell}$ in $R$ and taking the image of this fact under $g_{\ell}$ we have: $$R(g_{\ell}(h_{\ell}(\mathbf{x_0})), g_{\ell}(\mathbf{N_{\ell}})) = R(h''_{\ell}(\mathbf{x_0}), h''_{\ell}(\mathbf{y_0})).$$

The homomorphism $h''$ maps all the atoms of $\phi \wedge \psi$, in particular $R(\mathbf{x_0}, \mathbf{y_0})$ into facts in $db''_{\ell}$. Thus $g_{\ell}$ is a homomorphism. The only remaining thing is to show
\begin{equation*}
    \begin{split}
        & \forall i,j \in \mathbb{N}_0 \ \text{ such that }  i \neq j \\
        & \text{ and } g_i: db_i \mapsto db'_i \  \ \text{ and } g_j: db_j \mapsto db'_j , \\
        &\forall N \in \Null(db'_{\ell}) ~ ~ ~  g_{i}(N) \neq g_{j}(N) \\
    \end{split}
\end{equation*}
 For the nulls in $db'_{\ell}$ that are already in $db_{\ell}$ this is true because $h'$ is a homomorphism. A null $N'^{[s,e)}$ (replacing $y \in \mathbf{y}$) generated by the concrete chase step $K_c \xmapsto{\sigma^+, h} K'_c$ results in the labeled nulls $\langle N'_s, N'_{s+1}, \ldots, N'_{e-1} \rangle$ in the snapshots $db'_s , \ldots db'_{e-1}$ respectively. The homomorphism  $g'_{\ell}$ is an extension of $g_{\ell}$ such that 

\[g'_{\ell}(N) = 
\left\{
	\begin{array}{ll}
	h_s(N)  \mbox{ if } N \in \Null(db_s) \\ 
	h_{s+1}(N)  \mbox{ if } N \in \Null(db_{s+1}) \\ 
	\ldots \\
	h_{\ell}(N)  \mbox{ if } N \in \Null(db_{\ell}) \\
	\ldots \\
	h_{e-1}(N)   \mbox{ if } N \in \Null(db_{e-1}).\\
	\end{array}
\right.\]

{\em Case 2:} $\sigma^+$ is an egd. In this case the difference between $K_c$ and $K'_c$ is that some interval-annotated nulls in $K_c$ are replaced with other interval-annotated nulls or constants. But there is no new constant or interval-annotated null generated in $K'_c$.

If $\sigma^+$ can be applied on $K_c$ with $h$, based on Theorem \ref{thm:normprop} we know there is a homomorphism $h_{\ell}$ from $\phi(x)$ in $db_{\ell} \in \sem{K_c}$. 
As in case 1:
$$h'_{\ell} \circ h_{\ell}: \phi(x) \mapsto db''_{\ell},\  \  \ell \in \mathbb{N}_0,\  \ db''_{\ell} \in K_a.$$
Since each snapshot $db''_{\ell}$ ($\ell \in \mathbb{N}_0$) in $K_a$ satisfies the egd $\sigma$, $$h'_{\ell}(h_{\ell}(x_1)) = h'_{\ell}(h_{\ell}(x_2)).$$
We take $g_{\ell}$ to be $h'_{\ell}$. We need to show $h'_{\ell}$ is still a homomorphism from $db'_{\ell}$ to $db''_{\ell}$. The only way that $h'_{\ell}$ can fail to be a homomorphism on $db'_{\ell}$ is if $h'_{\ell}$ maps $h_{\ell}(x_1)$ and $h_{\ell}(x_2)$ into two different constants or labeled nulls of $db''_{\ell}$, which is not the case because $h'_{\ell}(h_{\ell}(x_1)) = h'_{\ell}(h_{\ell}(x_2)).$
\end{proof}
\begin{theorem}\label{thm:conunisol}
Assume a data exchange setting where $\Sigma^+_{st}$ consists of s-t tgds and $\Sigma^+_{eg}$ consists of egds.
\begin{enumerate}
    \item Let $(I_c, J_c)$ be the result some successful finite concrete chase of $(I_c, \emptyset)$ with $\Sigma^+_{st} \cup \Sigma^+_{eg}$. Then $\sem{J_c}$ is a universal solution.
    \item If there exists some failing chase of $(I, \emptyset)$ with $\Sigma^+_{st} \cup \Sigma^+_{eg}$, then there is no solution.
\end{enumerate}
   
\end{theorem}
\begin{proof}
{\em part 1: } 
The proof is based on Lemma \ref{lem: cchasestep} and the proof of Theorem 3.3 in \cite{FKMP05}. Let $J_a$ be an arbitrary solution (for example the result of chase on $\sem{I_c}$). Then $(\sem{I_c}, J_a)$ satisfies $\Sigma_{st} \cup \Sigma_{eg}$. The identity mapping $id : (\sem{I_c}, \emptyset) \mapsto (\sem{I_c}, J_a)$ is a homomorphism. By applying lemma \ref{lem: cchasestep} at each s-t tgd c-chase steps, we have $h_1: (\sem{I_c}, \sem{J'_c}) \rightarrow (\sem{I_c}, J_a)$. Then by applying lemma \ref{lem: cchasestep} at each egd chase step, we have $h: (\sem{I_c}, \sem{J_c}) \rightarrow (\sem{I_c}, J_a)$. In particular $h$ is a homomorphism from $\sem{J_c}$ to $J_a$. Thus, $\sem{J_c}$ is a universal solution.

{\em part 2: }Let $(I_c, J_c) \xmapsto{\sigma^+, h} \emptyset$ be the last egd c-chase step of a failing c-chase. Then $\sigma^+$ must be an egd in $\Sigma^+_{eg}$, say $\phi(\mathbf{x},t) \mapsto (x_1 = x_2)$ and $h: \phi(\mathbf{x},t) \mapsto J_c$ is a homomorphism such that $h(x_1)$ and $h(x_2)$ are {\em two distinct constants $a_1$ and respectively, $a_2$}. Suppose $h(t) = [s,e)$.  Let $\sem{J_c} = \langle db_0, db_1, \ldots, \rangle$. As in proof of Lemma \ref{lem: cchasestep} we have a homomorphism $h_{\ell}$ from lhs $\sigma: \phi(\mathbf{{x}}) \rightarrow x_1 = x_2$ to $db_\ell$ ($s \leq \ell < e$):
$$h_{\ell} : \phi(\mathbf{x}) \mapsto db_{\ell}$$

Assume by contradiction that there exists a solution $J_a = \langle db'_0, db'_1, \ldots, \rangle$. Since $J_a$ is a solution, $db'_{\ell} \models \sigma$, ($s \leq \ell <e$). The identity homomorphism 
$$id: (\sem{I_c}, \emptyset) \mapsto (\sem{I_c}, J_a) $$
implies, by Lemma \ref{lem: cchasestep}, the existence of homomorphism $g: (\sem{I_c}, \sem{J_c}) \mapsto (\sem{I_c}, J_a)$. In particular, $g$ is also a homomorphism from $\sem{J_c}$ to $J_a$, which means:
$$g_{\ell}:\  db_{\ell} \mapsto db'_{\ell} , \ \ \ell \in \mathbb{N}_0$$
Then $$g_{\ell} \circ h_{\ell}: \phi(\mathbf{x}) \mapsto db''_{\ell} \ \ s \leq \ell < e$$
Since $db'_{\ell} \models \sigma$, it must be the case that $g_{\ell} (h_{\ell} (x_1)) = g_{\ell} (h(x_2))$ and thus $g_{\ell} (a_1) = g_{\ell} (a_2)$. 
 Homomorphisms are identity on $\Const$, and so $a_1 = a_2$, which is a contradiction.
\end{proof}
\begin{corollary}\label{cor: homoequi}
Assume a data exchange setting in which $\Sigma^+_{st}$ consists of s-t tgds and $\Sigma^+_{eg}$ consists of egds. Let $I_c$ be a normalized concrete source instance w.r.t. $\Sigma_{st}$. Let $J_c$ be the result of c-chase on $I_c$. Let $J_a$ be the result of chase on $\sem{I_c}$. Then $\sem{J_c}$ is homomorphically equivalent to $J_a$, that is $\sem{J_c} \sim J_a$.
\end{corollary}

\section{Query Answering}\label{sec:QA}
In addition to finding a universal solution for a data exchange problem, another important issue in data exchange is query answering over the target schema~\cite{FKMP05}.
When queries are posed over the target schema,
different answers may be obtained depending on the solution that is considered.
To cope with the multiplicity of query results the notion of certain answers is used, where the
answers are the intersection of all the answers to the query on each possible solution \cite{FKMP05}.

Let $q$ be a non-temporal $k$-ary query, for $k\geq 0$.
Let $I_a$ be an abstract instance, that is $I_a= \langle db_0, db_1, \ldots \rangle$. Let $\mathcal{M}$ be a data exchange setting. The certain answers of $q$ w.r.t. $I_a$ and $\mathcal{M}$, denoted by $certain(q, I_a, \mathcal{M})$, is the sequence of sets of certain answers of $q$ on each snapshot
$$certain(q, I_a, \mathcal{M}) = \langle certain(q, db_0, \mathcal{M}), certain(q, db_1, \mathcal{M}), \ldots \rangle$$
where $certain(q, db_{\ell}, \mathcal{M})$ ( $\ell \in \mathbb{N}_0$) is the set of $k$-tuples $r$ of constants from $ db_{\ell}$,  such that for every solution $db'_{\ell}$ of the snapshot $db_{\ell}$ w.r.t. a schema mapping $\mathcal{M}$, $r \in q(db'_{\ell})$:  $$certain(q, db, \mathcal{M}) =\bigcap_{db' \text{ is a solution for } db w.r.t. \mathcal{M}}  q(db') $$
 
Na\"ive evaluation \cite{Abiteboul95, Arenas14, Imielinski86, FKMP05} is a technique commonly used in the literature to find certain answers for unions of conjunctive queries on na\"ive tables. 
It has been shown~\cite{Arenas14, FKMP05} that na\"ive evaluation of unions of conjunctive queries on a universal solution $db'$ for a relational source instance $db$ gives certain answers. Denote by $q(db)_{\downarrow}$ the result of na\"{i}ve evaluation of query $q$ on $db$ which is obtained by treating the labeled nulls as new constants (that is $N=N$, $N \neq M$, and $N \neq a$, where $a \in \Const$). It is shown that $certain(q, db, \mathcal{M}) = q(db')_{\downarrow}$ where $db'$ is a universal solution for db w.r.t. a data exchange setting ~\cite{Arenas14, FKMP05}. Denote by $q(J_a)_{\downarrow}$ the result of na\"{i}ve evaluation of the query $q$ on a universal solution $J_a = \langle db'_0, db'_1 , \ldots \rangle$ for a source instance $I_a$ w.r.t. a data exchange setting. Thus,
$$ceratin(q, I_a, \mathcal{M}) = q(J_a)_{\downarrow} = \langle q(db'_0)_{\downarrow}, q(db'_1)_{\downarrow}, \ldots \rangle.$$

Let $q$ be a query on the target schema in the abstract view. Denote by $q^+$ the corresponding query (obtained by augmenting all the atoms in the query $q$ by a free variable $t$) for the target schema in the concrete view.  
Given a union of conjunctive queries $q^+$ and a concrete solution $J_c$ for a source instance $I_c$ w.r.t. a data exchange setting, the na\"ive evaluation of $q^+$ on $J_c$, denoted by $q^+(J_c)_\downarrow$ is:
$$q^+(J_c)_\downarrow = \bigcup_{q'\text{ is a disjunct of } q^+} q'(J_c)_\downarrow$$
where $q'(J_c)_\downarrow$ is defined as follows:
\begin{enumerate}

\item Normalize instance $J_c$ w.r.t. $q'$. We denote the normalized instance by $J'_c$
\item Each interval-annotated null $N^{[s,e)}$ in $J_c$ is replaced with a fresh constant $cn^{[s,e)}$ everywhere it occurs. The result of this step is $J''_c$. 
\item Query $q$ is evaluated by finding all homomorphisms from variables in $q$ to $J''_c$. In particular, the variable $t$ is mapped to a time interval. The result of this step is denoted by $q(J''_c)$.
\item Tuples with fresh constants are dropped from $q(J''_c)$ to yield $q'(J_c)_\downarrow$.
\end{enumerate}

The following theorem shows that na\"ive evaluation on a concrete solution produces the same answers as na\"ive evaluation on the corresponding abstract solution under the semantic mapping.

\begin{theorem}\label{thm:con-abs-naive}
Let $J_c$ be a concrete solution for a source instance $I_c$ w.r.t. $\mathcal{M}$. Let $q^+$ be a union of conjunctive queries over the concrete target schema and $q$ the corresponding union of conjunctive queries on abstract target schema. Then
$\sem{q^+(J_c)_\downarrow} = q(\sem{J_c})_\downarrow$.
\end{theorem}
\begin{proof}
Let $\sem{J_c} = \langle db_0, db_1, \ldots \rangle$.
Let $(a_1, \ldots, a_k, [s,e))$ be a $(k+1)$-ary tuple in $q^+(J_c)_\downarrow$. Then, $\sem{q^+(J_c)_\downarrow} = \langle \mathbf{r_0}, \mathbf{r_1}, \ldots \rangle$ where $\mathbf{r_{\ell}}$ ( $\ell \in \mathbb{N}_0$) is a set of $k$-ary tuples defined as follows:
$$\mathbf{r_\ell} = \{\ (a_1, \ldots, a_k)\  | \ \exists i. \exists j.  \ (a_1, \ldots a_k, [i,j)) \in q^+(J_c)_\downarrow, \ i \leq \ell < j \}$$

Since $(a_1, \ldots, a_k, [s,e)) \in q^+(J_c)_\downarrow$ there is a homomorphism $h$ from a conjunctive query $q'$ that is a disjunct of $q^+$ to $J_c$.
W.l.o.g. we assume $q' : \exists y \phi^+ (\mathbf{x}, y, t)$. The proof can be easily extended when there is more than one existentially quantified variable.
Also, the snapshots $\mathbf{r_s}$ to $\mathbf{r_{e-1}}$ in $\sem{q^+(J_c)}$ contain the tuple $h(\mathbf{x})$ by definition.

Let $R^+(\mathbf{x}, y, t)$ be an arbitrary atom in $\phi^+$. Then $R^+(h(\mathbf{x}), h(y), h(t))$ is a fact in $J_c$. Depending on whether $h(y)$ is a constant or an interval-annotated null we consider two cases:

{\em Case 1: $h(y) = a^*$} where $a^* \in \Const$. In this case the snapshots $db_s$ to $db_{e-1}$ in $\sem{J_c}$ contains the fact $R(a_1, \ldots, a_k , a^*)$. Define homomorphisms $h_s, \ldots h_{e-1}$ as follows: $h_{\ell}(\mathbf{x}) = h(\mathbf{x})$ and $h_{\ell}(y) = h(y)$, $s \leq \ell < e$. Then $h_s, \ldots, h_{e-1}$ are homomorphisms from $
\phi(\mathbf{x},y)$ to $db_s, \ldots, db_{e-1}$, respectively because $R(h_{\ell}(\mathbf{x}), h_{\ell}(y))$ is a fact in $db_{\ell}$, $s \leq \ell < e$.
Hence, $h_{\ell}(\mathbf{x}) = (a_1, \ldots, a_k)$ is in $q(db_{\ell})$.
 
{\em Case 2: $h(y)= N^{[s,e)}$}. In this case, define  homomorphisms $h_s, \ldots h_{e-1}$ as follows: 
$h_{\ell}(\mathbf{x}) = h(\mathbf{x})$ and $h_{\ell}(y)= \pi_{\ell}(h(y)) = N_{\ell}$. Then $h_s, \ldots, h_{e-1}$ are homomorphisms from $
\phi(\mathbf{x},y)$ to $db_s, \ldots, db_{e-1}$, respectively. Therefore, the tuple $(a_1, \ldots, a_k)$ is in $q'(db_{\ell})$ and consequently in $q^+(J_c)_\downarrow$.

For the other direction, let $(a_1, \ldots, a_k)$ be a tuple in the result of $q$ on consecutive snapshots $db_s, db_{s+1}, \ldots db_{e-1}$, that is
$$(a_1, \ldots, a_k) \in q(db_{\ell}),\  s \leq \ell <e$$
Therefore, there exists a conjunctive query $\exists y \phi(\mathbf{x}, y)$ that is a disjunct of $q$ and there exists a homomorphism $h_{\ell}$ 
$$h_{\ell}:\  \phi(\mathbf{x}, y) \mapsto db_{\ell}.$$

That means if $R(\mathbf{x}, y)$ is an atom in $\phi$ then $R(h_{\ell}(\mathbf{x}), h_{\ell}(y))$ is in $db_{\ell}$. Observe that in this direction we also assume one existentially variable $y$ in the conjunctive query. But this assumption is without loss of any generality. 

Based on the value of $h_{\ell}(y)$ we consider two cases:

{\em Case 1: $h_{\ell}(y)= a^*$}, where $a^* \in \Const$. 
Thus, $R^+(a_1, \ldots, a_k, a^*, [s,e))$ is a fact in $J_c$.
Define a homomorphism $h$ on variables in $\phi^+(\mathbf{x}, y, t)$ (in a disjunct in $q^+$) as follows:

\[h(z) = 
\left\{
	\begin{array}{ll}
	h_{\ell}(z) & \mbox{ if } h_{\ell}(z) \mbox{ is a constant }\\ 
	{[s,e)} & \mbox{ if } z=t\\
	\end{array}
\right.\]

$h$ is a homomorphism from $\phi^+(\mathbf{x}, y, t)$ to $J_c$ because it maps an arbitrary atom such as $R^+(\mathbf{x}, y , t)$ in $\phi^+$ to a fact $R^+(h(\mathbf{x}), h(y), h(t))$ in $J_c$. Therefore, $(a_1, \ldots a_k, [s,e))$ is in $q^+(J_c)_{\downarrow}$ which means $\mathbf{r_{\ell}}$, $s \leq \ell < e$ contains the tuple $(a_1, \ldots a_k)$.

\noindent{\em Case 2: $h_{\ell}(y) = N_{\ell}$}, $s \leq \ell < e$. In this case by definition of $\sem{J_c}$, $R^+(a_1, \ldots, a_k, N^{[s,e)}, [s,e))$ is a fact in $J_c$.

\[h(z) = 
\left\{
	\begin{array}{lll}
	h_{\ell}(z) & \mbox{ if } h_{\ell}(z) \mbox{ is a constant }\\ 
	N^{[s,e)} &\mbox{ if } h_{\ell}(z)= N_{\ell} \\ 
	{[s,e)} & \mbox{ if } z=t\\
	\end{array}
\right.\]
Same as previous case, $h$ is a homomorphism from $\phi^+(\mathbf{x}, y, t)$ to $J_c$ and $\mathbf{r_{\ell}}$ in $\sem{q^+(J_c)}$ contains the tuple $(a_1, \ldots a_k)$.

\end{proof}
\begin{corollary}\label{cor:connaiv_crtnans}
Let $J_c$ be the result of concrete chase on a concrete source instance $I_c$ w.r.t. a temporal schema mapping $\mathcal{M}= (R_S, R_T, \Sigma_{st}, \Sigma_{eg})$.  
Let $q^+$ be a union of conjunctive queries over the target schema. Then  
$certain(q,\sem{I_c}, \mathcal{M}) = \sem{q^+(J_c)_\downarrow}$.
\end{corollary}

\section{Related Work}\label{sec:relatedwork}
Here we overview the previous relevant work on data exchange and temporal databases.
The formal foundations of data exchange were developed by Fagin et al. in~\cite{FKMP05}. The authors showed that the chase algorithm, previously used for checking implication of data dependencies, can be used to produce a universal solution for instances of the data exchange problem. Universal solutions map homomorphically to other solutions for the source instance. This property makes them the preferred solutions to query answering. Universal solutions and, in general, solutions of instances of data exchange problem can contain incomplete information. Representing incomplete information and evaluating queries over them are more complex than in the complete case as shown by Imielinski and Lipski~\cite{Imielinski86} as well as Abiteboul et al. in \cite{ Abiteboul95}. The gap between the theoretical work on incomplete information and what has been used in practice is discussed by Gheerbrant et al.~\cite{Gheerbrant14} and Libkin \cite{Libkin14}. The chase algorithm proposed by Fagin et al. in~\cite{FKMP05} produces labeled nulls for incomplete values. Relations containing labeled nulls are called {\em na\"ive tables}~\cite{Abiteboul95, Imielinski86}. In data exchange the semantics of answering queries is defined in terms of {\em certain answers}~\cite{Arenas14, FKMP05}. Certain answers \cite{Imielinski86} are tuples that belong to the answer of the posed query no matter which solution is used.
 Fagin et al. showed that whenever a universal solution can be computed in polynomial time (for the class of dependencies identified in~\cite{FKMP05}), certain answers to unions of conjunctive queries can also be computed in polynomial time in data complexity in~\cite{FKMP05}. Computing certain answers for queries that have more than one inequality, however, is a coNP-complete problem~\cite{FKMP05}. Data exchange and incomplete information and other possible semantics for query answering are discussed in detail by Libkin in \cite{Libkin06}.

The formal foundations of temporal data models and query languages were studied by Chomicki in~\cite{Chomicki94} and by Chomicki and Toman in ~\cite{Chomicki05}. Abstract versus concrete temporal views were first developed in the context of the semantics of temporal query languages~\cite{Toman96}. These kinds of views of temporal data were also used in program debugging and dynamic program analysis \cite{Lessa11}. However, Chomicki~\cite{Chomicki94} and Chomicki and Toman~\cite{Chomicki05} did not discuss incomplete temporal information and its possible semantics. The notion of normalization was previously used in the context of query answering in temporal databases~\cite{Lessa11, Toman97} on tuples with the same schema that agree on non-temporal attribute values. The normalization algorithms that we introduced change the concrete instance w.r.t. conjunctions of atomic formulas. Koubarakis proposed a unified framework for both finite and infinite, definite and indefinite temporal data ~\cite{Koubarakis94a, Koubarakis94b}.  His suggested framework extends {\em conditional tables} (a.k.a. c-tables) ~\cite{Imielinski86} and can be used to store facts such as roomA is booked from 2 to {\em sometime between 5 to 8}. He used global conditions to define the constraints on the start point or end point of a time interval. In his framework, the temporal attribute values can be unknown. C-tables are a generalization of na\"ive tables where a table is associated with global and local conditions specified by logic formulas~\cite{Imielinski86}. In Koubarakis framework the indefinite (incomplete) temporal data are not the result of data exchange. His framework, does not deal with schema mappings or integrity constraints. We proposed interval-annotated nulls for unknown values in concrete data exchange to align the semantics of temporal data exchange with the data exchange on abstract view. Also in our framework the value of the temporal attribute is known because the schema mappings are non-temporal. Therefore, there is no condition on the temporal attribute or non-temporal attributes as a result of data exchange. Na\"ive tables are sufficient for representing incomplete information in temporal data exchange with non-temporal schema mappings.

\section{Conclusion and Future work}\label{sec:conclusion}
In this paper, we proposed a framework for data exchange on temporal data which relies on the distinction between the abstract and concrete view of the data. Abstract view is responsible for the semantics while the concrete view is used in the implementations. We considered a basic case where the schema mapping is non-temporal. We first extended the standard chase procedure on abstract instances. Defining chase on the abstract instances provides a conceptual tool on how a concrete chase should work. Then we defined a concrete chase on concrete instances. We showed normalization of the concrete instance is necessary to define homomorphisms from the lhs of a dependency with a shared temporal variable among the atoms to a concrete instance. We finished the paper by showing the result of the concrete chase is a good candidate to be materialized and used for answering queries.

A natural extension of this paper is to enrich the schema mappings such that they can express temporal phenomena.  For example, temporal modal operators such as $\lozenge$ ({\em sometime in the future}), $\square$ ({\em always in the future}), $\blacklozenge$ ({\em} sometime in the past) and $\blacksquare$ ({\em always in the past}) can be added to the language. 
For example, consider the following constraint which says every PhD graduate was a PhD candidate at some point before they graduate and they had a topic and an adviser.  
$$\square(\forall n\  PhDgrad (n) \rightarrow \blacklozenge \exists adv, top \ PhDCan(n,adv,top))$$

This constraint is equivalent to the following constraint in {\em two-sorted FOL (2-FOL)}\cite{Chomicki05}:
$$\forall n, t ~PhDgrad(n,t) \rightarrow \exists ~adv,~top,~ t' \ PhDCan(n, adv, top, t') \wedge ~t' < t$$
At any snapshot $db_{\ell}$ if there is a fact about a PhD graduate in $PhDgrad$, then a snapshot $db_{i}$, $i<\ell$, should contain a fact about the PhD graduate in $PhDCan$ with a topic and an adviser. Considering these schema mappings, 
the notions of chase steps, solutions and universal solutions should be redefined. As an example if $\blacklozenge$ is used in the rhs of a dependency (such as the above dependency), is it enough to choose an arbitrary snapshot and generate facts according to the rhs of the dependency in that snapshot? What will be a universal solution in this case? 

The schema mappings can also be enriched with linear order $<$ as in \cite{AfratiCK11} and with arithmetic operations as in \cite{tenCate13}. The linear order in \cite{AfratiCK11} is interpreted over an arbitrary countable
dense linear order without endpoints while the domain of time points that we consider in this dissertation is discrete linear order. Afrati et al. \cite{AfratiCK11} conjecture   
that the results they obtained, regarding data exchange and query answering in presence of arithmetic operations, would change significantly in discrete ordered domains.

Another direction of research is to revisit the classical data exchange problems in the context of temporal databases such as the notion of {\em core}~\cite{FaginKP05} and comparing open world assumption and closed world assumption \cite{HernichLS11,Libkin09}.  
\newpage
\noindent\textbf{Acknowledgement}

We would like to thank Wang-Chiew Tan for her participation in the initial phases of this project. This research is supported by NSF awards IIS-1524469 and IIS-1450590.
\bibliography{arxivev2}

\end{document}